\documentclass{emulateapj}
\usepackage{apjfonts}

\usepackage{lscape}

\usepackage{graphicx}
\usepackage{natbib}
\newcommand{\hbeta}{H{$\beta$}}
\newcommand{\halpha}{H{$\alpha$}}

\newcommand{\OIIIa}{[O{\sevenrm\,III}]\,$\lambda$4959}
\newcommand{\OIIIb}{[O{\sevenrm\,III}]\,$\lambda$5007}

 \font\sevenrm=cmr7 scaled 1000

\begin{document}

\title{Identifying Supermassive Black Hole Binaries with Broad Emission Line Diagnosis}

\shorttitle{SPECTROSCOPIC SMBH BINARIES}

\shortauthors{SHEN \& LOEB}

\author{Yue Shen and Abraham Loeb\\
Harvard-Smithsonian Center for Astrophysics, 60 Garden Street,
MS-51, Cambridge, MA 02138, USA}


\begin{abstract}
Double-peaked broad emission lines in Active Galactic Nuclei (AGNs) may indicate the existence of a bound supermassive black hole (SMBH) binary where two distinct broad line regions (BLRs) contribute together to the line profile. An alternative interpretation is a disk emitter origin for the double-peaked line profile. Using simple BLR models, we calculate the expected broad line profile for a SMBH binary at different separations. Under reasonable assumptions that both BLRs are illuminated by the two active SMBHs and that the ionizing flux at the BLR location is roughly constant, we confirm the emergence of double-peaked features and radial velocity drifts of the two peaks due to the binary orbital motion. However, such a clear double-peaked feature only arises in a particular stage of the binary evolution when the two BHs are close enough such that the line-of-sight orbital velocity difference is larger than the FWHM of the individual broad components, while the two BLRs are still mostly distinct. Prior to this stage, the velocity splitting due to the orbit motion of the binary is too small to separate the emission from the two BLRs, leading to asymmetric broad line profiles in general. When the two BHs are even closer such that the two BLRs can no longer be distinct, the line
profile becomes more complex and the splitting of the peaks does not correspond to the orbital motion of the binary. In this regime there are no coherent radial
velocity drifts in the peaks with time. Asymmetric line profiles are probably a far more common signature of binary SMBHs than are double-peaked profiles. We discuss the
temporal variations of the broad line profile for binary SMBHs and
highlight the different behaviors of reverberation mapping in the
binary and disk emitter cases, which may serve as a feasible tool to disentangle these two scenarios.
\end{abstract}


\section{Introduction}

Binary supermassive black holes (SMBHs) are generic
outcomes of hierarchical galaxy mergers
\citep[e.g.,][and references therein]{Colpi_Dotti_2009}. A couple of galactic-scale binary AGNs were reported based on spatially resolved imaging and spectroscopy
\citep[e.g.,][]{Komossa_etal_2003,Bianchi_etal_2008,Comerford_etal_2009b}.
These binaries are at projected separations of the order of kpc,
below which it is difficult to spatially resolve both SMBHs at
cosmological distances. While in principle radio interferometry
can resolve close binaries down to millarcsecond resolution, it
requires both black holes (BHs) to be radio sources and so far
there is only one sub-kpc binary candidate detected in the radio
\citep[][]{Rodriguez_etal_2006}.

Characteristic velocity offsets or double-peaked features in AGN
emission line (both broad and narrow lines) profiles have been invoked to indicate the presence of a binary SMBH, even if its spatial extent is not resolved. While
this idea is not new
\citep[e.g.,][]{Heckman_etal_1981,Gaskell_1983,Peterson_etal_1987},
it only became feasible recently to search for such binary candidates in a systematic way using large statistical samples with good spectral
quality (most notably the Sloan Digital Sky
Survey\footnote{http://www.sdss.org/} (SDSS) samples).
Increasingly larger spectroscopic samples are
starting to offer a unique opportunity to search for candidate
binary SMBHs based on their spectral
properties
\citep[e.g.,][]{Zhou_etal_2004,Bonning_etal_2007,Komossa_etal_2008a,Comerford_etal_2009a,
Boroson_Lauer_2009,Liu_etal_2009,Smith_etal_2009,Shields_etal_2009,Wang_etal_2009a,
Xu_Komossa_2009}.
This is an important first step towards quantifying the frequency
of binary SMBHs at various separations, and providing constraints on hierarchical
galaxy merger models and predictions for future low-frequency
gravitational wave experiments such as the Laser Interferometer
Space Antenna\footnote{http://lisa.nasa.gov/} (LISA).

While it has become routine to select candidate binaries from
large spectroscopic data sets, these candidates are
less secure than spatially resolved cases. The usual
difficulty involves the poorly understood emission line region
geometry and kinematics even for single black holes (BHs), which may mimic a binary system. It is
rather difficult to rule out one or the other based on a
single-epoch spectrum alone. In the case of kpc separation binary SMBHs,
high spatial-resolution imaging and spectroscopy may potentially resolve both active BHs and therefore confirm the binary nature \citep[e.g,][]{Liu_etal_2010b}. For gravitationally bound binary SMBHs ($\la 10$ pc), spectral monitoring and reverberation mapping may disentangle the
binary scenario from its alternatives
\citep[e.g.,][]{Gaskell_1983,Gaskell_2009,Eracleous_etal_1997,Gezari_etal_2007}.
At even smaller separations, \citet{Loeb_2009} suggested that the time dependence of the broad
lines due to orbital motion can be detected in binaries on the
verge of entering the gravitational wave dominated inspiral.

Given the importance of binary SMBHs for galaxy formation models
and for future low-frequency gravitational wave detection
experiments, it is crucial to understand the dynamics and geometry
of emission line regions in a binary system and to identify such
binaries in a systematic way. However, despite of the ongoing efforts
\citep[e.g.,][]{Escala_etal_2005,Dotti_etal_2006,Dotti_etal_2007,Mayer_etal_2007,
Bogdanovic_etal_2008,Cuadra_etal_2009}, it is still
challenging to explore the parameter space in detail in numerical
simulations with the adequate dynamical range and the necessary
input physics. On the other hand, observational constraints on the dynamics and geometry of AGN emission line regions have not yet converged to provide reliable inputs for numerical simulations. Here we use simple toy models to study the feasibility of using
broad line diagnostics to identify {\em bound} binary SMBHs, taking into account both dynamical and ionization effects of the two BHs on the combined BLRs. In \S\ref{sec:model} we develop toy models to predict broad line profiles under various circumstances. In \S\ref{sec:temporal_variation} we discuss the temporal properties
of the AGN spectrum in different scenarios. Finally, we discuss
our results in \S\ref{sec:disc}. Unless otherwise stated, we will use H$\beta$ as the fiducial emission line since this is the best-studied broad line in reverberation mapping studies, from which characteristic BLR properties are best determined.


\section{broad emission line profiles}\label{sec:model}

\begin{figure}
  \centering
    \includegraphics[width=0.48\textwidth]{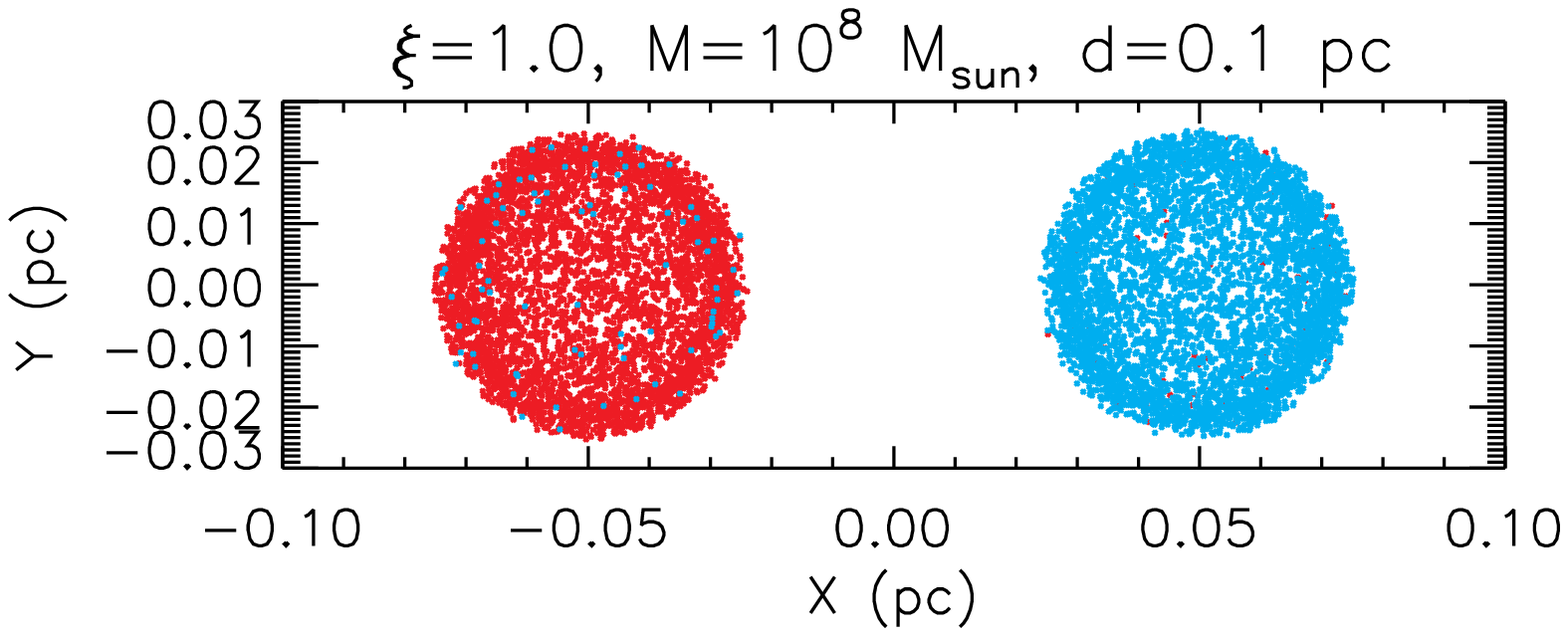}
    \includegraphics[width=0.45\textwidth]{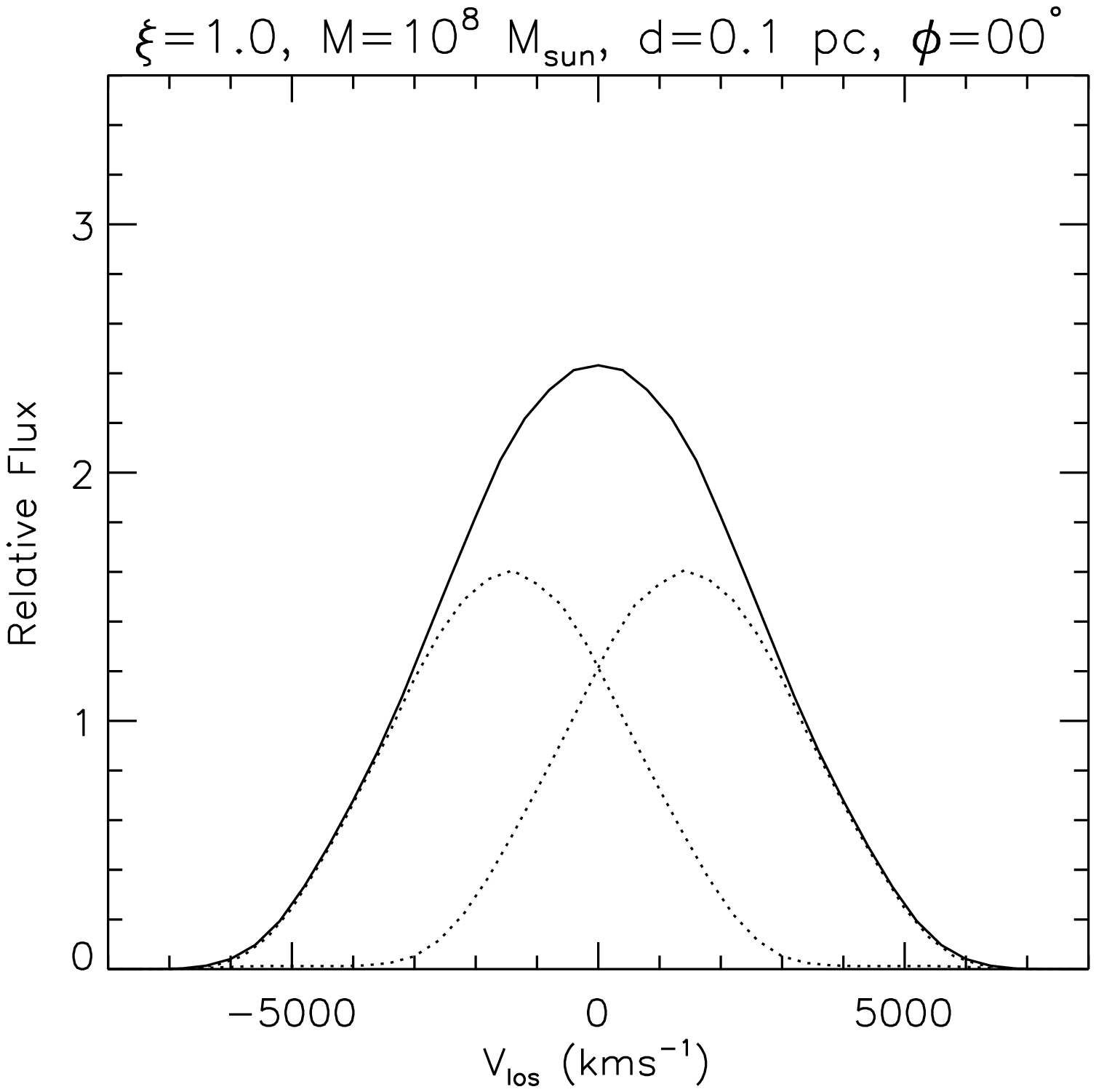}
    \includegraphics[width=0.45\textwidth]{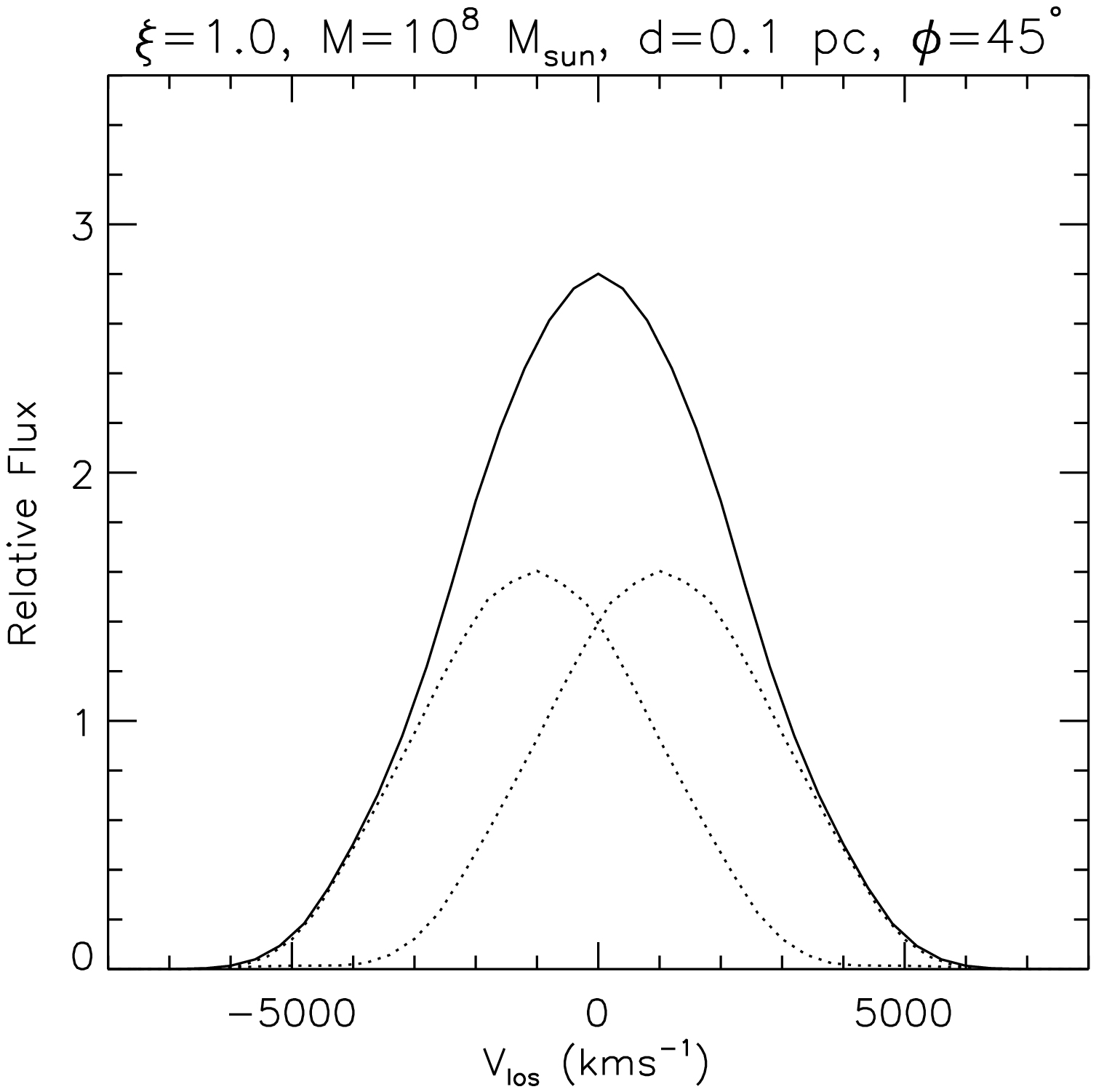}
    \caption{{\em Top}: Distributions of BLR clouds (projected onto the binary orbital plane) for a binary of two $10^8\ M_\odot$ BHs with a separation
     $d=0.1$ pc. For clarity we only show a small fraction of randomly selected test
     particles. Different colors indicate BLR clouds initially associated with the two BHs.
     The rotation of the binary is counterclockwise. The observer is located in the $xy$ plane at $y=+\infty$, and the radial velocities of the two BHs are maximal at this phase.
     The few clouds that become mixed with the other BH were initially on highly eccentric orbits with large apocenter or on hyperbolic orbits which travel to the vicinity of
     the other BH later. We did not remove such clouds in our simulation as they have essentially no effect on the derived line profile.
     {\em Middle}: Line profile when the radial velocities of the two BHs are maximal (orbital phase
     angle $\phi=0^\circ$). The dotted lines are individual contributions from the two BHs. {\em Bottom}:
     Line profile after $1/8$ of the orbital period ($\phi=45^\circ$).}
    \label{fig:M1d8_d0d1}
\end{figure}

\begin{figure}
  \centering
    \includegraphics[width=0.48\textwidth]{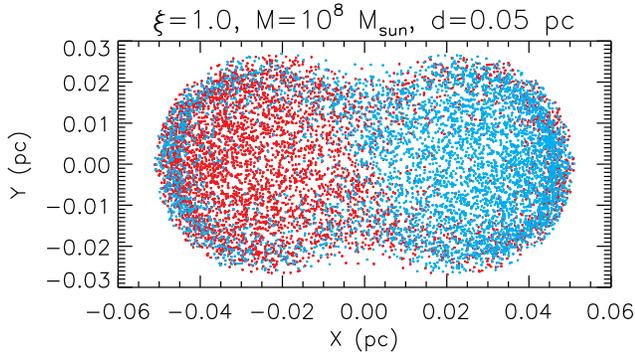}
    \includegraphics[width=0.45\textwidth]{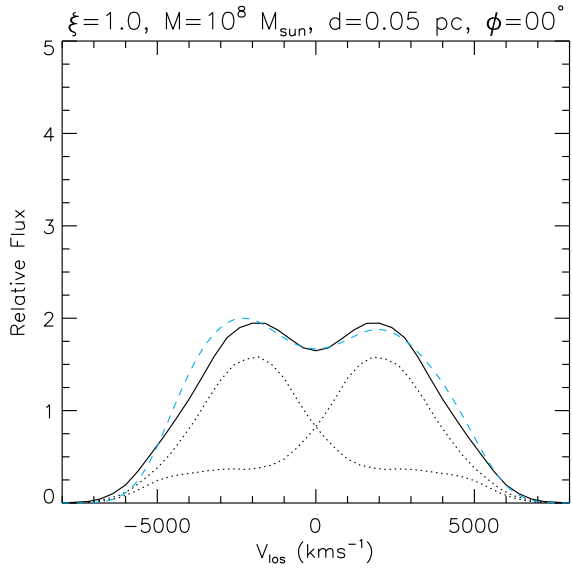}
    \includegraphics[width=0.45\textwidth]{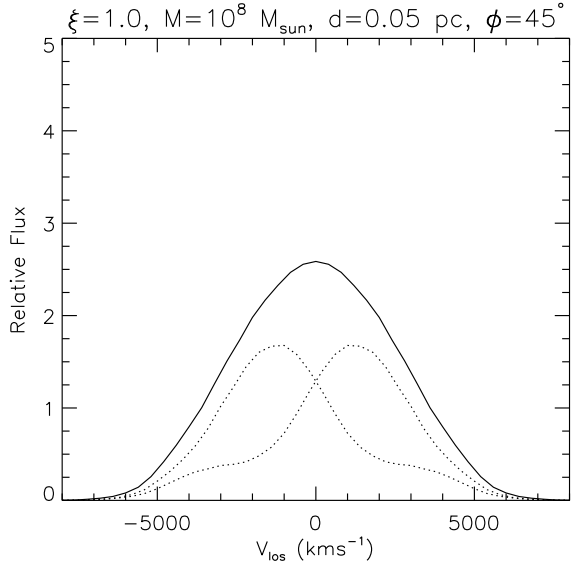}
    \caption{Distributions of BLR clouds (projected onto the binary orbital plane) and line profiles for a binary of two $10^8\ M_\odot$ BHs with a separation
     $d=0.05$ pc. Notations are the same as in Fig.\ \ref{fig:M1d8_d0d1}. The cyan dashed line in the middle
     panel shows a disk emitter model (see the text for details).}
    \label{fig:M1d8_d0d05}
\end{figure}

\begin{figure}
  \centering
    \includegraphics[width=0.48\textwidth]{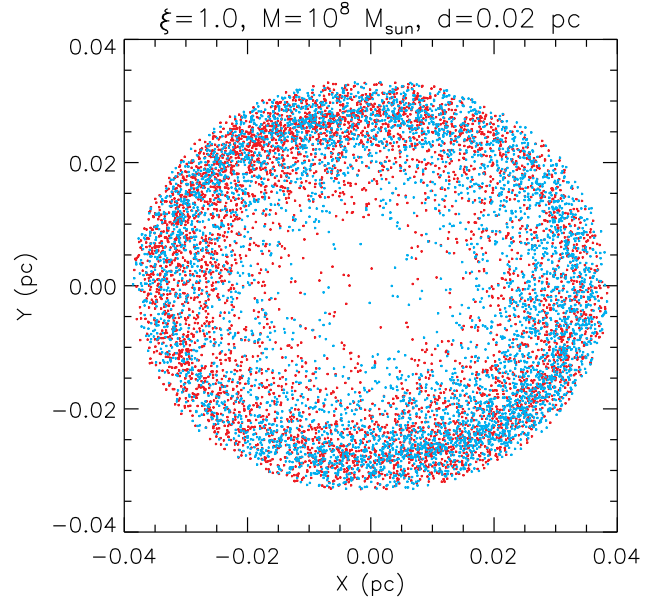}
    \includegraphics[width=0.45\textwidth]{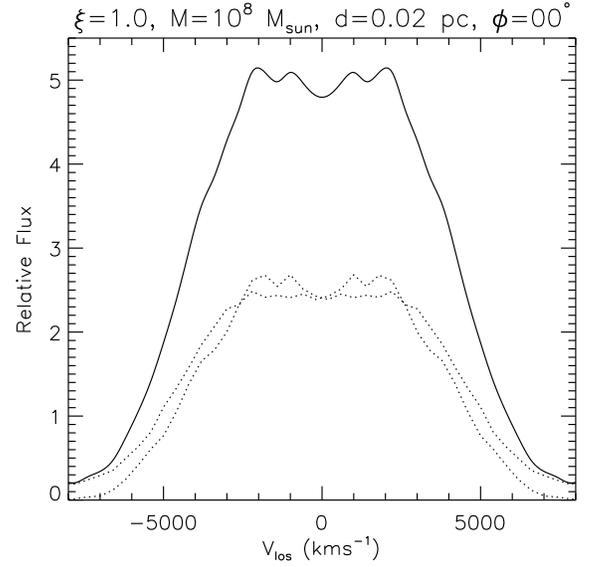}
    \includegraphics[width=0.45\textwidth]{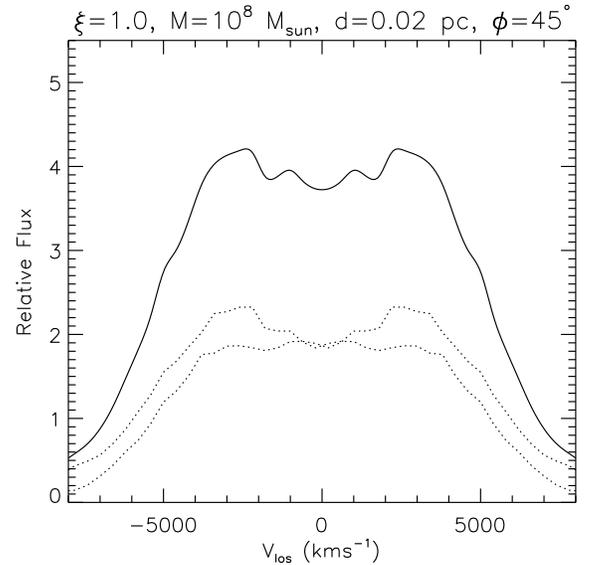}
    \caption{Distributions of BLR clouds (projected onto the binary orbital plane) and line profiles for a binary of two $10^8\ M_\odot$ BHs with a separation
     $d=0.02$ pc. Notations are the same as in Fig.\ \ref{fig:M1d8_d0d1}.}
    \label{fig:M1d8_d0d02}
\end{figure}


We consider the broad line profiles for bound binary SMBHs where the dynamics is dominated by the gravitational potential of the two BHs,
assuming both BHs are active and have their own BLRs. We also assume that the two BHs are corotating on a circular orbit. We are
interested in binary SMBHs with comparable masses ($0.3\le
\xi\equiv M_1/M_2 \le 1$), since binaries with extreme mass ratios
are either difficult to detect with broad line diagnosis (if line emission is proportional to BH mass), or difficult to form due to the extended
dynamical friction time of the galaxy merger. For simplicity, we
also assume a fixed Eddington ratio $\lambda_{\rm Edd}=0.1$ \citep[e.g.,][]{Shen_etal_2008b}, but
we note that in practice the two active black holes could have
different Eddington ratios. For demonstration purposes we will
only show binary examples with an edge-on view (Figs.\
\ref{fig:M1d8_d0d1}-\ref{fig:temp_vari_inter}). The effect of
inclination $I$ is to reduce the radial velocities by $\sin I$. We
use $\phi$ to denote the binary orbital phase with $\phi=0^\circ$
when the radial velocities of the two BHs are maximal.

If both BHs have their own {\em distinct} BLR, and each BLR is
corotating along with its BH in the binary orbit, we
expect to see two time-varying velocity components in the broad
line profile. The velocity splitting of the two components
depends on the binary orbit separation and the BLR sizes.
When the orbital separation is large, the velocity splitting of
the two broad line components is small compared with the broad line width and the two components will blend with each other in the spectrum. When the two BHs come closer, the
velocity splitting gets wider, and a double-peak profile may
emerge. When the two BHs come even closer, both BLRs are
dynamically influenced by the two BHs, and so some BLR material becomes circumbinary, leading to a more complex velocity
structure in the combined line profile. The velocity peaks in the broad line, if any, will not simply correspond to the orbital motion of the binary in this case. Eventually, the two BHs will get so
close that they dynamically affect the BLR clouds like a single
BH, and the broad line profile may become single-peaked again. These simple arguments suggest that a clear double-peaked broad line profile may only arise during a particular stage of the binary orbit evolution, where the orbital velocity of the binary is large enough to split the two peaks and both BLRs are still mostly distinct.

Denoting the separation of the binary as $d$, and the BLR radius in a single BH
as $R_{\rm BLR}$, the criterion for both BLRs to be bound to their
own BHs is $R_{\rm BLR}\la 0.5d$. For a single BH, assuming the
BLR is photoionized by the continuum luminosity from the central BH, the $R_{\rm BLR}-L$ relation (assuming $R_{\rm BLR}\propto L^{1/2}$)
derived from reverberation mapping
\citep[e.g.,][]{Kaspi_etal_2000,Peterson_etal_2004,Mclure_Jarvis_2002,Kaspi_etal_2007}
gives
\begin{equation}\label{eqn:BLR_size}
R_{\rm BLR}\approx 2.2\times 10^{-2}\bigg(\frac{\lambda_{\rm
Edd}}{0.1}\bigg)^{1/2}\bigg(\frac{M_{\bullet}}{10^8\
M_\odot}\bigg)^{1/2}\ {\rm pc}\ ,
\end{equation}
where $M_\bullet$ is the BH mass and $\lambda_{\rm Edd}$ is the
Eddington ratio of the broad line AGN\footnote{The $R_{\rm BLR}-L$
relation is usually calibrated using the restframe 5100 \AA\
continuum luminosity. We have adopted a bolometric correction
${\rm BC}_{5100\,{\textrm \AA}}=10$ to convert the continuum
luminosity to bolometric luminosity.}. This $R_{\rm BLR}-L$ scaling implies that the ionization parameter,
$U\propto L/r^2$, is roughly constant for a particular line species (assuming constant electron density). The orbital period for a
circular orbit at this location is:
\begin{equation}\label{eqn:t_orb}
t_{\rm orb}=30\ \bigg(\frac{\lambda_{\rm
Edd}}{0.1}\bigg)^{3/4}\bigg(\frac{M_\bullet}{10^8\,
M_\odot}\bigg)^{1/4}\ {\rm yr}\ .
\end{equation}

Assuming the BLR is virialized, the FWHM of a single broad
component is:
\begin{eqnarray}\label{eqn:FWHM}
v_{\rm FWHM}&=&\bigg(\frac{GM_{\bullet}}{fR_{\rm
BLR}}\bigg)^{1/2}\nonumber\\
&\approx& 4200 \bigg(\frac{\lambda_{\rm
Edd}}{0.1}\bigg)^{-1/4}\nonumber\\
&\times&\bigg(\frac{M_\bullet}{10^8M_\odot}\bigg)^{1/4}\bigg(\frac{f}{1.4}\bigg)^{-1/2}\
{\rm km\,s^{-1}}\ ,\nonumber\\
\end{eqnarray}
where $f$ is the virial coefficient accounting for our ignorance
of the BLR geometry, and is of order of unity. A recent
determination of $f$ based on the reverberation BH masses and BH
masses from the $M_\bullet-\sigma$ relation gives $f\approx 1.4$
\citep[][]{Onken_etal_2004}\footnote{Note that we have used a different definition of the
virial coefficient from others. The coefficient $f$ here corresponds to the coefficient $\epsilon$ in \citet[][]{Onken_etal_2004}.}. 

If the two BLRs are distinct and simply corotate along with their own BHs in a binary system, the line-of-sight (LOS) velocity splitting of the two broad line peaks changes periodically in time $t$:
\begin{eqnarray}\label{eqn:v_split}
v_{\rm los}&=&\bigg[\frac{GM_{\rm tot}}{d}\bigg]^{1/2}\sin I
\sin(2\pi t/P)\ ,\nonumber\\
&=& 6300\bigg(\frac{M_{\rm tot}}{10^8\,
M_\odot}\bigg)^{1/2}\bigg(\frac{d}{0.01\ {\rm
pc}}\bigg)^{-1/2}\nonumber\\
&\times&\sin I\sin(2\pi t/P)\ {\rm
km\,s^{-1}}\ ,\nonumber\\
\end{eqnarray}
where $M_{\rm tot}=M_1+M_2$ is the total mass of the two BHs, $I$
is the inclination of the binary orbital plane relative to the
LOS, and
\begin{eqnarray}\label{eqn:bin_p}
P&\equiv& 2\pi d^{3/2}(GM_{\rm tot})^{-1/2}\nonumber\\
&=&9.5\times 10^{3}\bigg(\frac{d}{1\,{\rm
pc}}\bigg)^{3/2}\bigg(\frac{M_{\rm tot}}{10^8\,
M_\odot}\bigg)^{-1/2}\ {\rm yr}
\end{eqnarray}
is the orbital period of the binary.

From these simple calculations
(\ref{eqn:BLR_size})-(\ref{eqn:v_split}) we can investigate how
well-separated the double components could be without violating
the assumption that each BLR is mostly under the influence of only one
BH. In order to see a clear double-peaked
feature we require $v_{\rm los,max}\ga v_{\rm FWHM}$, while in
order to have distinct BLRs we need $R_{\rm BLR}\la d/2$. Assuming
$\xi\equiv M_1/M_2\le 1$ these two criteria imply:
\begin{eqnarray}\label{eqn:window}
0.044\bigg(\frac{M_2}{10^8\ M_\odot}\bigg)^{1/2}{\rm\ pc}\la d\la
0.063\bigg(\frac{1+\xi}{2}\bigg)\bigg(\frac{M_2}{10^8\
M_\odot}\bigg)^{1/2}{\rm\ pc}\
.\nonumber\\
\end{eqnarray}
Therefore, there is only a narrow window of separation (as well as mass ratio $\xi$) within which a double-peaked broad line profile may emerge
without violating the assumption that both BLRs are distinct.

\subsection{Model Setup}\label{sec:model}

Next, we investigate the broad line profile for a binary SMBH in a more quantitative manner. First we describe our approach to model the BLRs of a binary SMBH.

We start with a simple prescription for the BLR around a single BH, where the BLR is assumed to be an assembly of discrete clouds \citep[e.g.,][]{Peterson_1997}. Then two BHs with their individual BLRs are placed on a circular orbit. We integrate numerically the orbits of individual clouds (treated as test particles) in the circular restricted three body problem. By imposing an ionization condition (see below) we identify the clouds that will radiate the line emission and determine the line profile based on the LOS velocity distribution of these line-emitting clouds. This procedure is detailed below.

First, we need to specify a model for the BLR around a single BH. Despite decades of research, the detailed structure of BLR is still poorly constrained. The most powerful observational tool to study BLR structure is reverberation mapping. But even with the best-studied reverberation mapping sample, there is still no complete consensus on the general BLR structure \citep[e.g.,][]{Denney_etal_2009}. Nevertheless, reverberation mapping does provide a characteristic scale for the BLR radius, given by the $R_{\rm BLR}-L$ relation in Eqn.\ (\ref{eqn:BLR_size}). In a handful of cases, reverberation mapping of different line species in the same system shows that lines with larger ionization parameters have smaller radii from the central BH \citep[e.g.,][]{Peterson_Wandel_2000}, and that the velocity inferred from the line width is consistent with virialized motion.

Motivated by these observations, we adopt the following simple model for the BLR around a single BH. For a single BH, we uniformly populate clouds as test particles within a spherical shell with inner and outer
radii $\sqrt{0.8}$ and $\sqrt{1.2}$ of the radius in Eq.\
(\ref{eqn:BLR_size}), i.e., the flux or ionization parameter (assuming constant electron density) required to photoionize a particular line species in the BLR
clouds is roughly constant within $\pm 20\%$ (the exact value of the percentage does not change our conclusions). The velocities of
those clouds are generated from a Maxwellian distribution whose $1$D
dispersion is determined from the virial relation,
$\sigma=\sqrt{GM_\bullet/(3r)}$, with random orientations. Since it is unphysical to restrict these clouds within a perfectly thin shell, we
evolve the Keplerian orbits of these clouds in the single BH
system using standard analytical formulae \citep[e.g.][]{Murray_Dermott_1999}. The system quickly establishes a quasi-steady state configuration where the spatial and velocity distributions converge after $\sim t_{\rm orb}$ (Eqn.\ \ref{eqn:t_orb}). The resulting quasi-steady-state distributions are taken as the initial cloud
distributions around a single BH\footnote{We do not remove the
small fraction ($\sim 10\%$) of clouds on hyperbolic orbits in the initial distributions, since such clouds will either be bound in the binary
system, or will escape to large distances and not contribute to line emission anyway. We verified that removing these clouds does not have any effect on the derived line profiles.}. These clouds have a radial distribution peaked around
the initial shell location but also extend to larger and smaller radii, and the Gaussianity of their velocity distribution is mostly preserved. We generate 5000 test particles for each BH. As discussed later, the sporadic snapshots of these test particle trajectories will be combined to compute the cloud distributions in a binary system. The resulting line profiles are very smooth even though the total number of test particles used is modest.

This simple model for the BLR around a single BH is highly idealized, given the limited understandings of actual BLR properties. It nevertheless reproduces line profiles and characteristic BLR sizes that are consistent with observations. We have assumed that only clouds at a narrow range of distances from the BH can be line-emitting and constitute the BLR. In reality the BLR in some AGNs could have a larger spatial extent, as inferred in some (albeit limited) reverberation mapping studies \citep[e.g.,][and references therein]{Denney_etal_2009,Horne_etal_2004}. This can be understood if the electron density in the BLR varies with radius, so that the flux ($\propto r^{-2}$) required to produce the proper ionization parameter also has a larger range in photoionization models. However, given the small scatter in the mean $R_{\rm BLR}-L$ relation \citep[e.g.,][]{Peterson_2010}, our fiducial choice of a relatively thin shell geometry for the BLR should be a reasonable approximation at least for the majority of the line-emitting clouds in the BLR. One caveat is that we are not including a significant fraction of clouds that are either far more distant or closer in than the characteristic BLR distance. Such clouds will not be line-emitting in the single BH system because the ionization parameter is either too high or too low. Clouds much closer in will be tightly bound to their BH and will not contribute to the line emission in the binary system either. However, if there are a large amount of cold clouds orbiting outside the BLR in single BHs, these clouds are easier to become circumbinary and may be ionized by the combined continua from both BHs in the binary system, thus making the double-peaked feature less prominent. Below we proceed with our fiducial model for the BLR around single BHs, and the effects of such a distant cloud reservoir will be further discussed in Sec \ref{subsec:disc1}.

We combine the two BH plus clouds systems to
form a binary on a circular orbit. In the frame corotating with
the binary, we derive the instantaneous locations and velocities
of these clouds by orbital integrations of the restricted three-body
problem. We are only interested in the temporal behaviors over a period that can be monitored on human life timescales, so we integrate the system for a few hundred years sampled with 1,000 evenly distributed temporal snapshots. Long-term stability of the cloud orbits are not considered in this paper\footnote{It is possible that some of these orbits are unstable over longer timescales and the long-lived orbits may have somewhat different spatial and velocity distributions. On the other hand, the individual BLRs are likely non-static and some mechanisms may exist to replenish the clouds continuously (such as a wind from the accretion disk). It is beyond the scope of this paper to take into account these complications, but we note that our integration time may be too short for some of the cases studied here.}. We use a Bulirsch-Stoer integrator \citep[][]{NR} and the relative accuracy tracked by the Jacobi constant was well below $10^{-10}$ over the course of the integration. We then combine all
the snapshots as a quasi steady-state configuration of clouds in
the binary system, so as to improve the statistics of the cloud distributions. We found that the cloud distributions (in real and velocity space) in individual snapshots are similar (but with much poorer statistics) to those from the combined snapshots, after the initial $\sim 10\%$ of integration time, i.e., the velocity distribution of BLR clouds quickly establishes a quasi-equilibrium in much less than one binary orbit time. Although clouds are still constantly expelled by the
binary afterwards, the loss of clouds does not have significant effects on the velocity structure of the BLR over the course of the integration.

Once we have the locations and velocities of the clouds, we determine line-emitting BLR clouds under the
assumption that clouds are illuminated by both BHs, and that the combined
flux (or ionization parameter) required to ionize a BLR cloud is roughly constant (within
$\pm 20\%$, assuming a constant electron density). Quantitatively, the locations of these line-emitting clouds satisfy the following constraint:
\begin{eqnarray}
\frac{0.8}{(2.2\times 10^{-2})^2}
&\le&\frac{\displaystyle\big(\frac{\lambda_{\rm
Edd,1}}{0.1}\big)\big(\frac{M_{1}}{10^8\ M_\odot}\big)}{(r_1/{\rm
pc})^2}+\frac{\displaystyle\big(\frac{\lambda_{\rm
Edd,2}}{0.1}\big)\big(\frac{M_{2}}{10^8\ M_\odot}\big)}{(r_2/{\rm
pc})^2}\nonumber\\
&\le& \frac{1.2}{(2.2\times 10^{-2})^2}\ ,
\end{eqnarray}
where $r_1$ and $r_2$ are the distances to the two BHs. This
condition follows the photoionization criterion for single BHs in
Eq.\ (\ref{eqn:BLR_size}).


Based on the procedure outlined above, we derive the broad line
profile for binary SMBHs at any orbital phase and hence the
temporal variation of line profile due to the orbital motion of
the binary.

Figures \ref{fig:M1d8_d0d1}-\ref{fig:M1d8_d0d02} show several
examples for an equal-mass binary with $M_1=M_2=10^8\ M_\odot$,
and separations of $d=0.1,0.05,0.02$ pc, with integration time 2, 6 and 22 times the binary orbital period given by Eq.\ (\ref{eqn:bin_p}), respectively. Only the intermediate separation case ($d=0.05$ pc) satisfies the constraint of Eq.\ (\ref{eqn:window}). As the two BHs come closer, their BLRs become less distinct, and eventually the two BHs start to affect both BLRs in terms of their
dynamics and illumination. In the example with the closest
separation (e.g., Fig.\ \ref{fig:M1d8_d0d02}), there is
practically a single BLR surrounding the two BHs, and the velocity
structure of these BLR clouds are more complex than in the
distinct BLR cases. Many BLR clouds are now on circumbinary orbits (such as horseshoe or tadpole-type orbits around the
triangular Lagrangian points) and there are no coherent
radial velocity shifts in the two peaks due to the orbital motion
of the binary.

\begin{figure}
  \centering
    \includegraphics[width=0.48\textwidth]{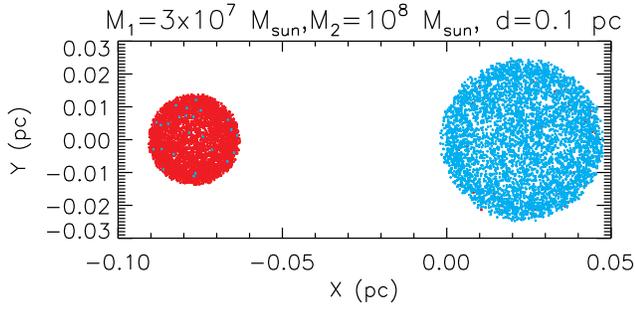}
    \includegraphics[width=0.45\textwidth]{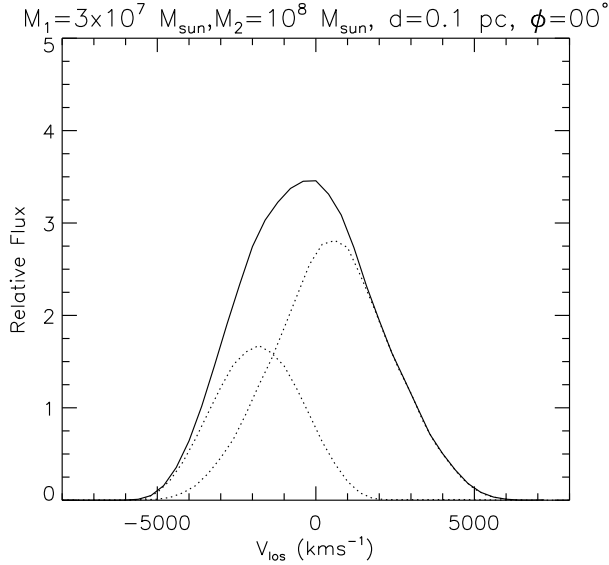}
    \includegraphics[width=0.45\textwidth]{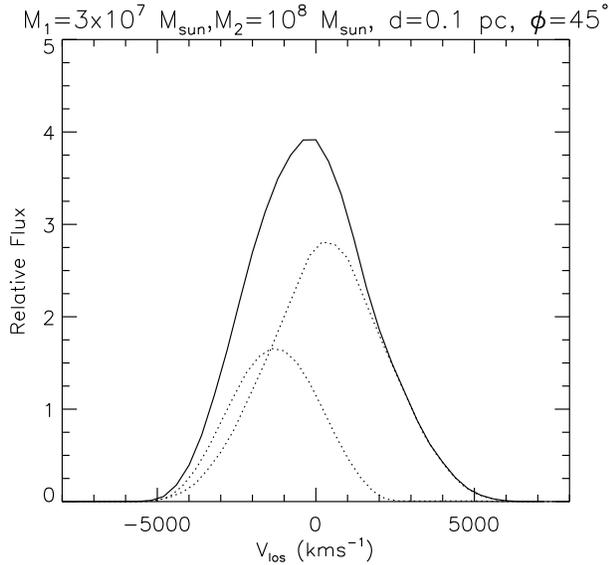}
    \caption{Distributions of BLR clouds (projected onto the binary orbital plane) and line profiles for a SMBH binary with $M_1=3\times 10^7\ M_\odot$ and $M_2=10^8\ M_\odot$ with a separation
     $d=0.1$ pc. Notations are the same as in Fig.\ \ref{fig:M1d8_d0d1}.}
    \label{fig:M3d7_M1d8_d0d1}
\end{figure}

\begin{figure}
  \centering
    \includegraphics[width=0.48\textwidth]{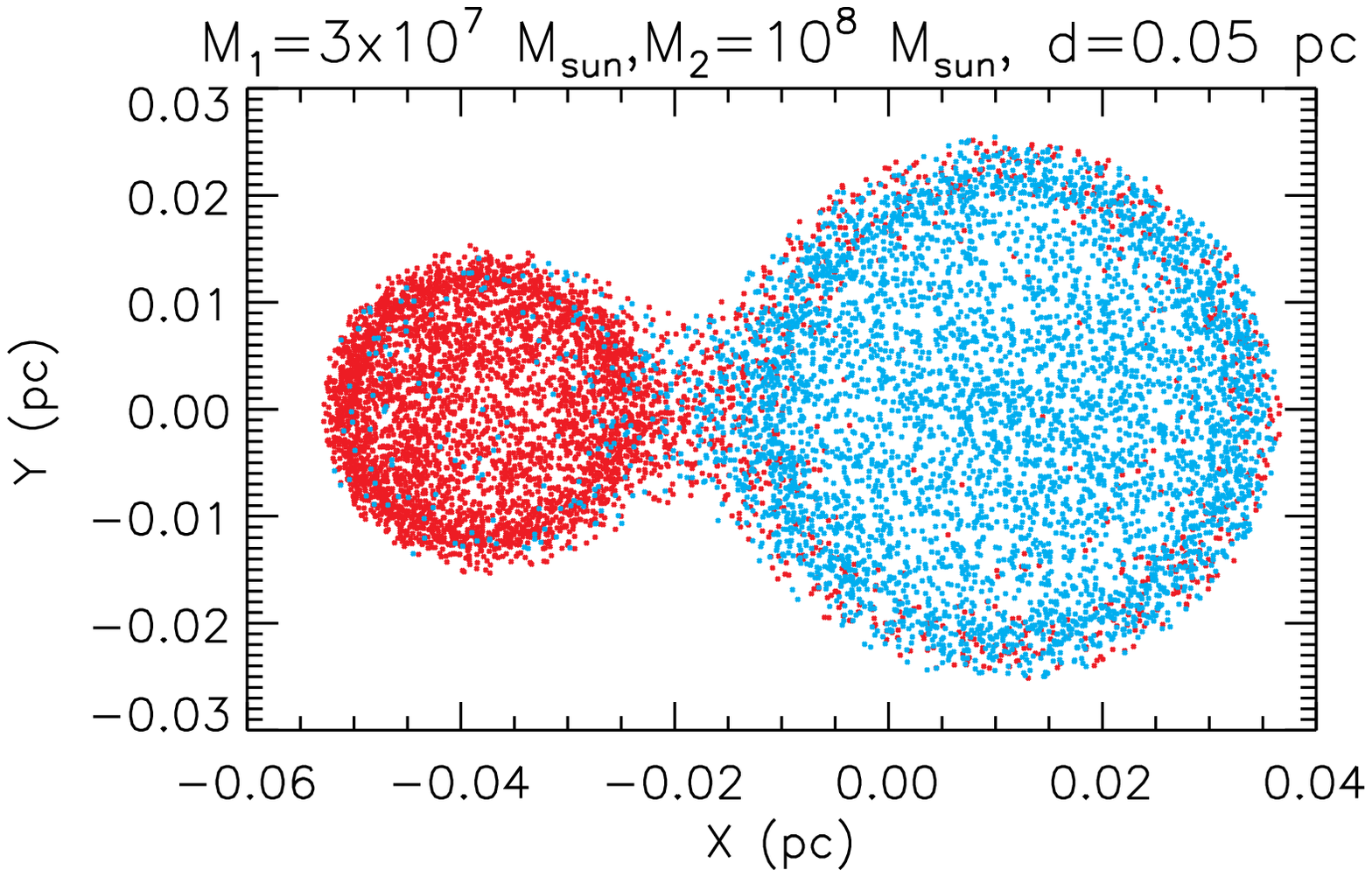}
    \includegraphics[width=0.45\textwidth]{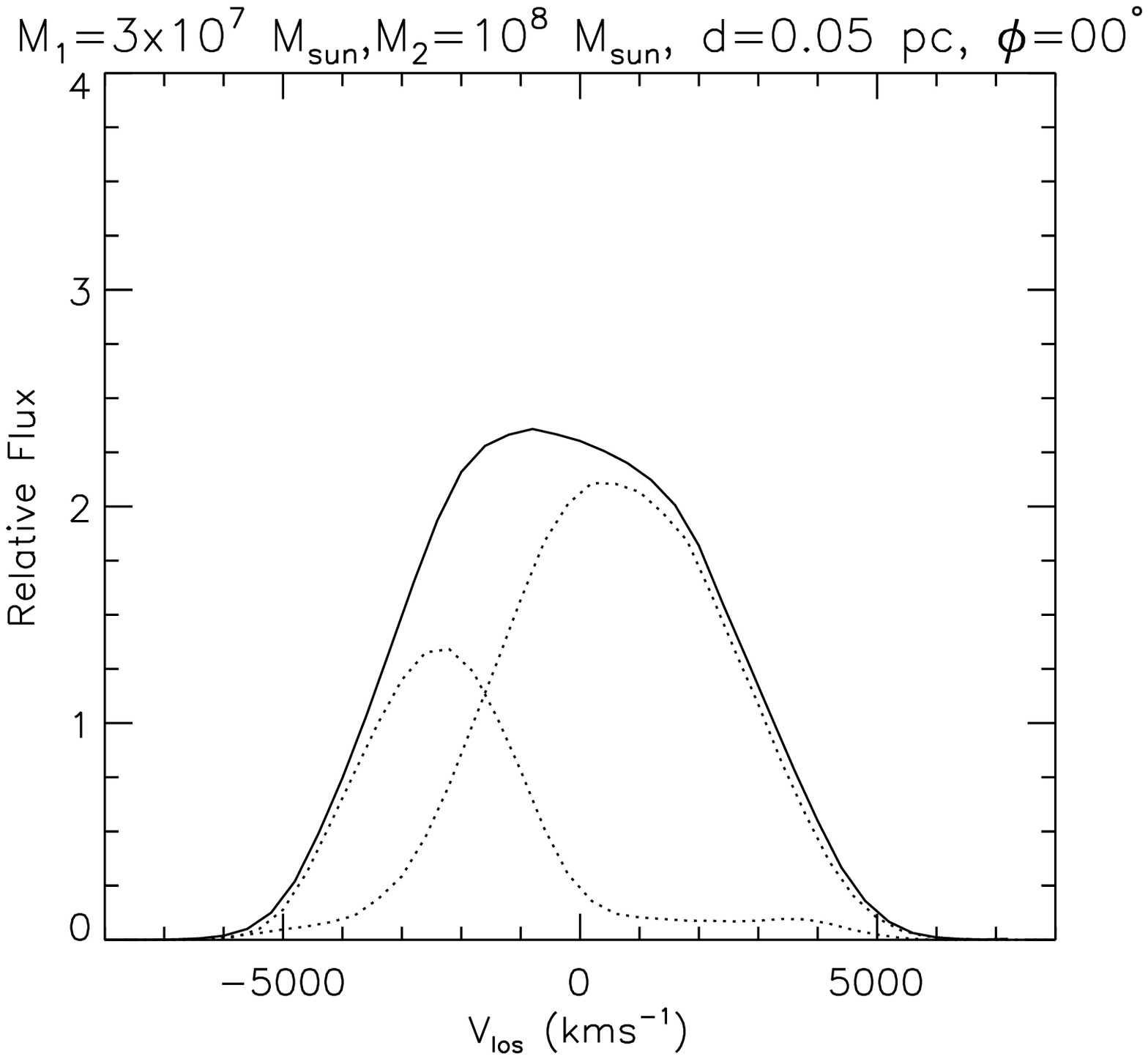}
    \includegraphics[width=0.45\textwidth]{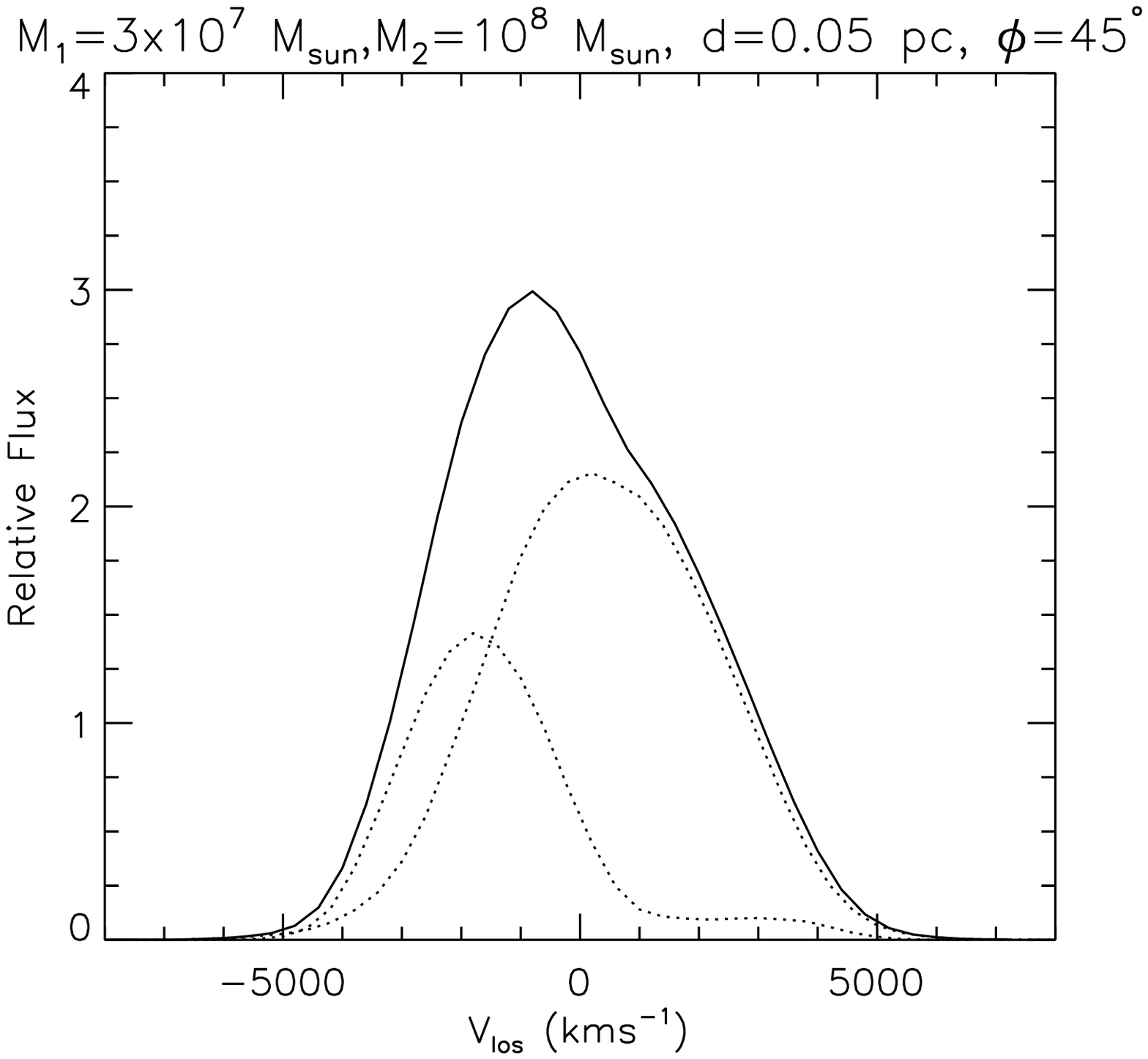}
    \caption{Distributions of BLR clouds (projected onto the binary orbital plane) and line profiles for a SMBH binary with $M_1=3\times 10^7\ M_\odot$ and $M_2=10^8\ M_\odot$ with a separation
     $d=0.05$ pc. Notations are the same as in Fig.\ \ref{fig:M1d8_d0d1}.}
    \label{fig:M3d7_M1d8_d0d05}
\end{figure}

\begin{figure}
  \centering
    \includegraphics[width=0.48\textwidth]{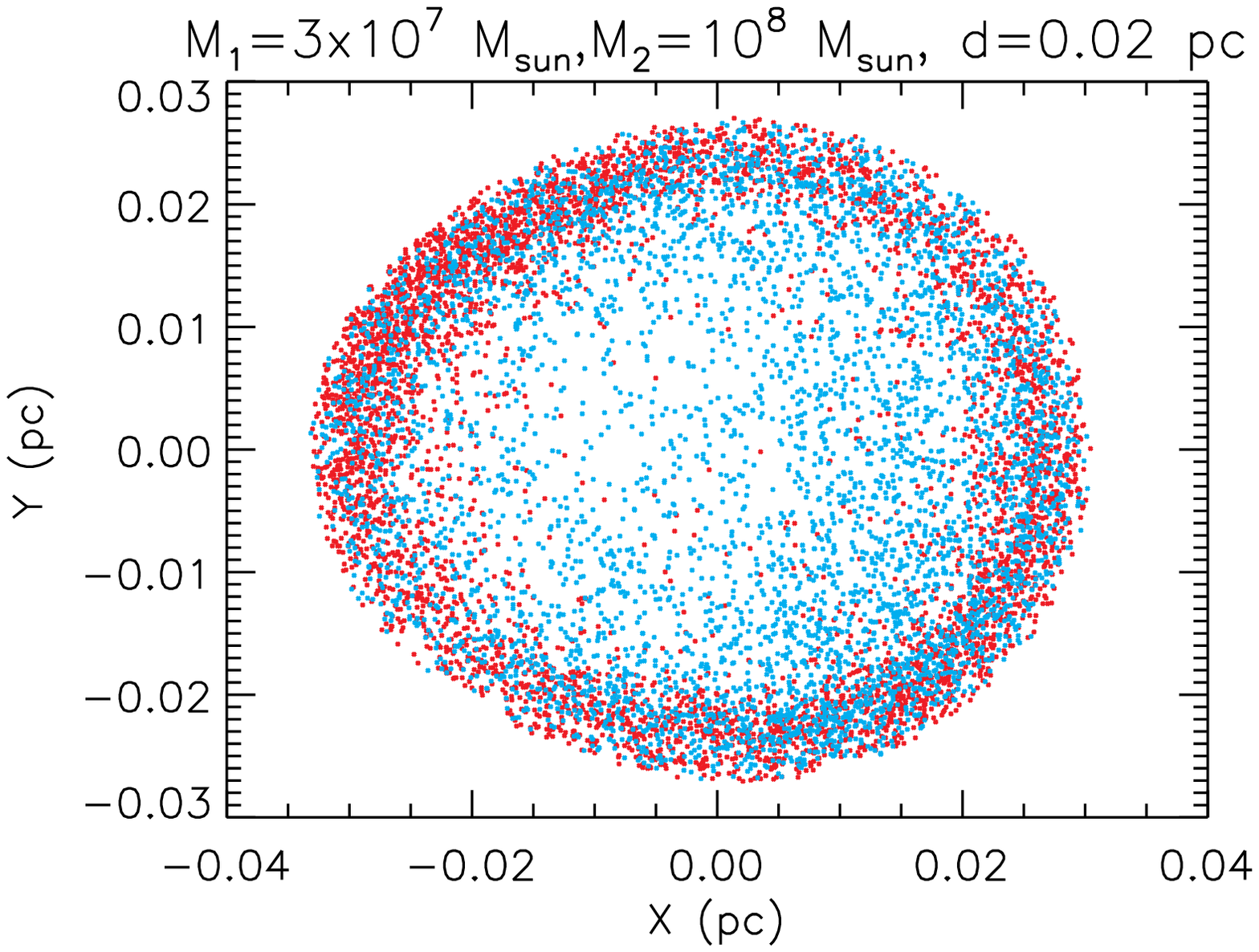}
    \includegraphics[width=0.45\textwidth]{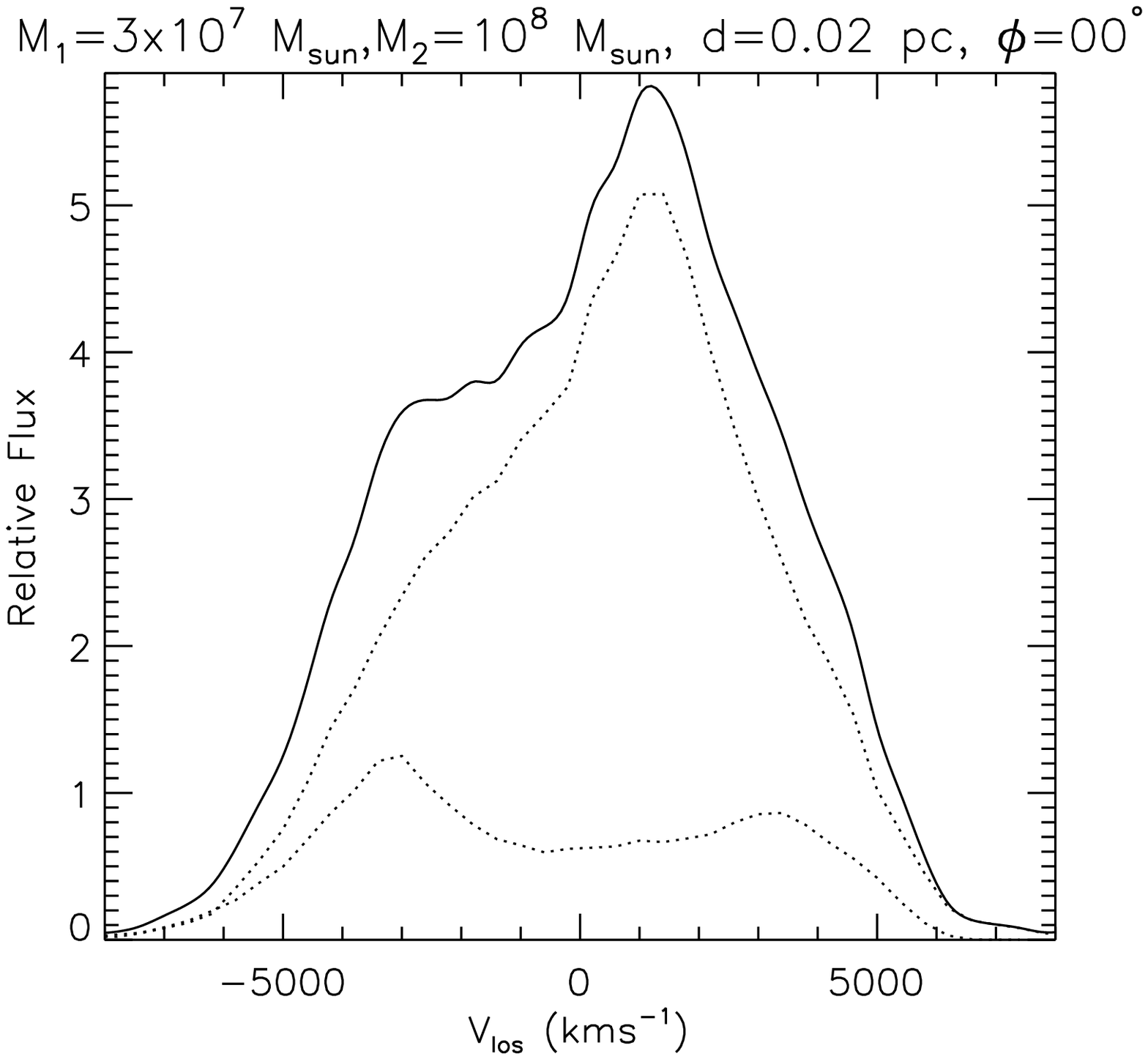}
    \includegraphics[width=0.45\textwidth]{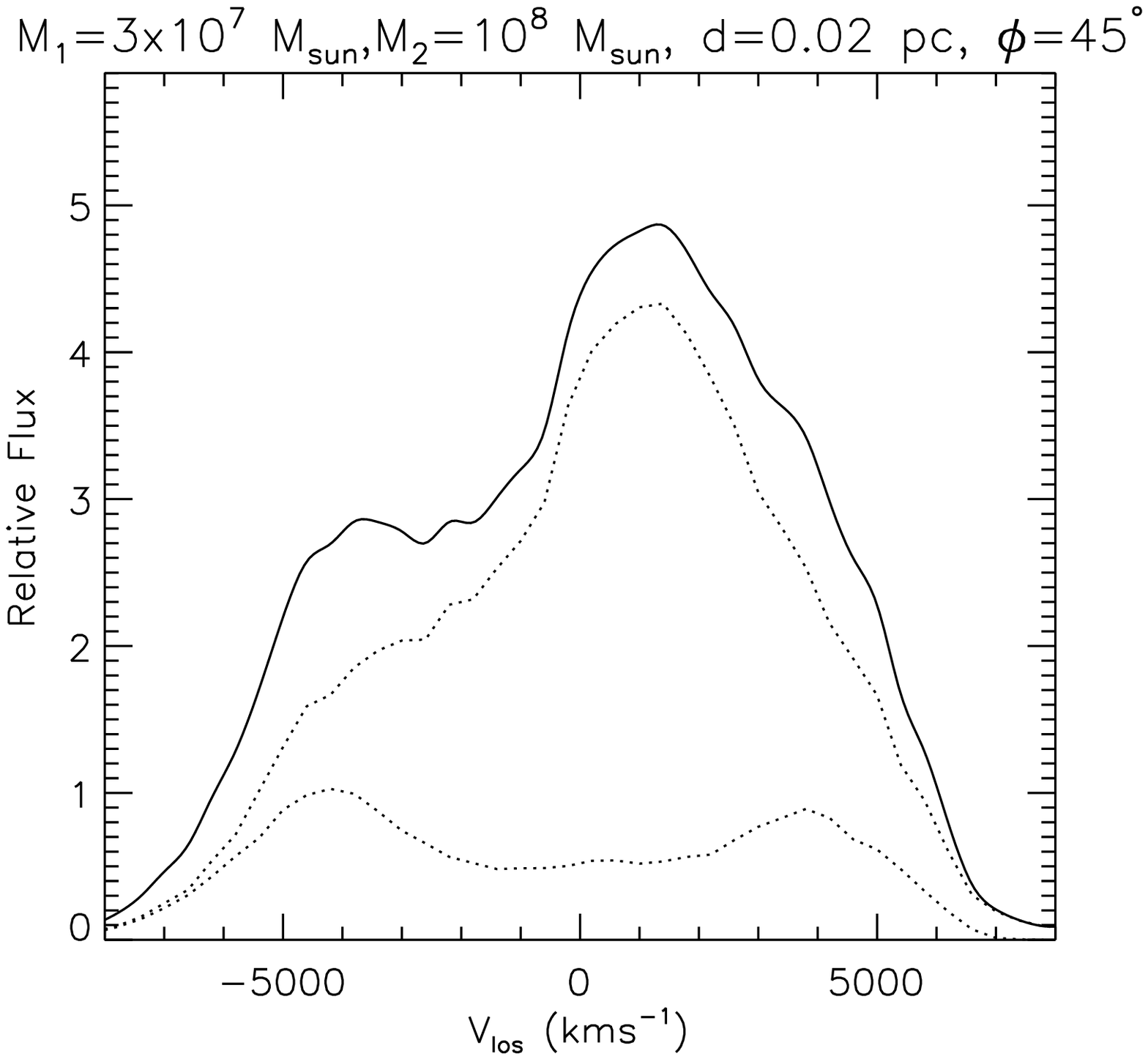}
    \caption{Distributions of BLR clouds (projected onto the binary orbital plane) and line profiles for a SMBH binary with $M_1=3\times 10^7\ M_\odot$ and $M_2=10^8\ M_\odot$ with a separation
     $d=0.02$ pc. Notations are the same as in Fig.\ \ref{fig:M1d8_d0d1}.}
    \label{fig:M3d7_M1d8_d0d02}
\end{figure}

Figures \ref{fig:M3d7_M1d8_d0d1}-\ref{fig:M3d7_M1d8_d0d02} show
several examples for a non-equal-mass binary with
$M_1=3\times10^7\ M_\odot$ and $M_2=10^8\ M_\odot$ and separations
of $d=0.1,0.05,0.02$ pc, where the BLR clouds associated with each
individual BH are given a flux weight proportional to the BH mass. The integration times are 2, 6 and 22 times the binary orbital period, respectively. Again, only the intermediate separation case ($d=0.05$ pc) satisfies the constraint of Eq.\ (\ref{eqn:window}).
The qualitative difference from the equal binary case is the
increase in line asymmetry and decrease in the prominence of the
double peaks \citep[see][for some similar line profiles generated with different model prescriptions for the binary BLRs]{Popovic_etal_2000}.

\subsection{Disk Emitters}

An alternative interpretation of double-peaked broad line profiles
is the disk emitter scenario, where the anomalous broad line
emission originates from a relativistic accretion disk around a
single BH
\citep[][]{Chen_Halpern_1989,Eracleous_Halpern_1994,Eracleous_etal_1995}.
In this case the blueshifted and redshifted components originate
from the part of the disk moving towards and away from the
observer. The disk emitter model has been successful in
reproducing the line profile in many double-peaked broad line AGNs
\citep[e.g.,][]{Chen_Halpern_1989,Eracleous_Halpern_1994,Eracleous_etal_1995,
Eracleous_etal_1997,Strateva_etal_2003,Luo_etal_2009}.
As an example, in the middle panel of Fig.\ \ref{fig:M1d8_d0d05}
we show a disk emitter model for an inclined (inclination
$I=25^\circ$) elliptical disk (with eccentricity $e=0.1$) with
inner and outer radii of $500r_g$ and $5000r_g$ (where
$r_g=GM/c^2$ is the gravitational radius), specific intensity
$I_\nu\propto r^{-3}$, an internal turbulent broadening
$\sigma=800\ {\rm kms^{-1}}$ and a major axis orientation of the
elliptical disk $\phi_0=60^\circ$ \citep[][]{Eracleous_etal_1995},
around a BH with $M=10^8\ M_\odot$. The disk emitter model clearly
shows a double-peaked profile, resembling that for a SMBH binary.
But the temporal variations of the broad line in the disk emitter
scenario are different from those for the binary SMBH scenario, as
we discuss next.

\section{Temporal Variations}\label{sec:temporal_variation}


\begin{figure*}
  \centering
    \includegraphics[width=0.45\textwidth]{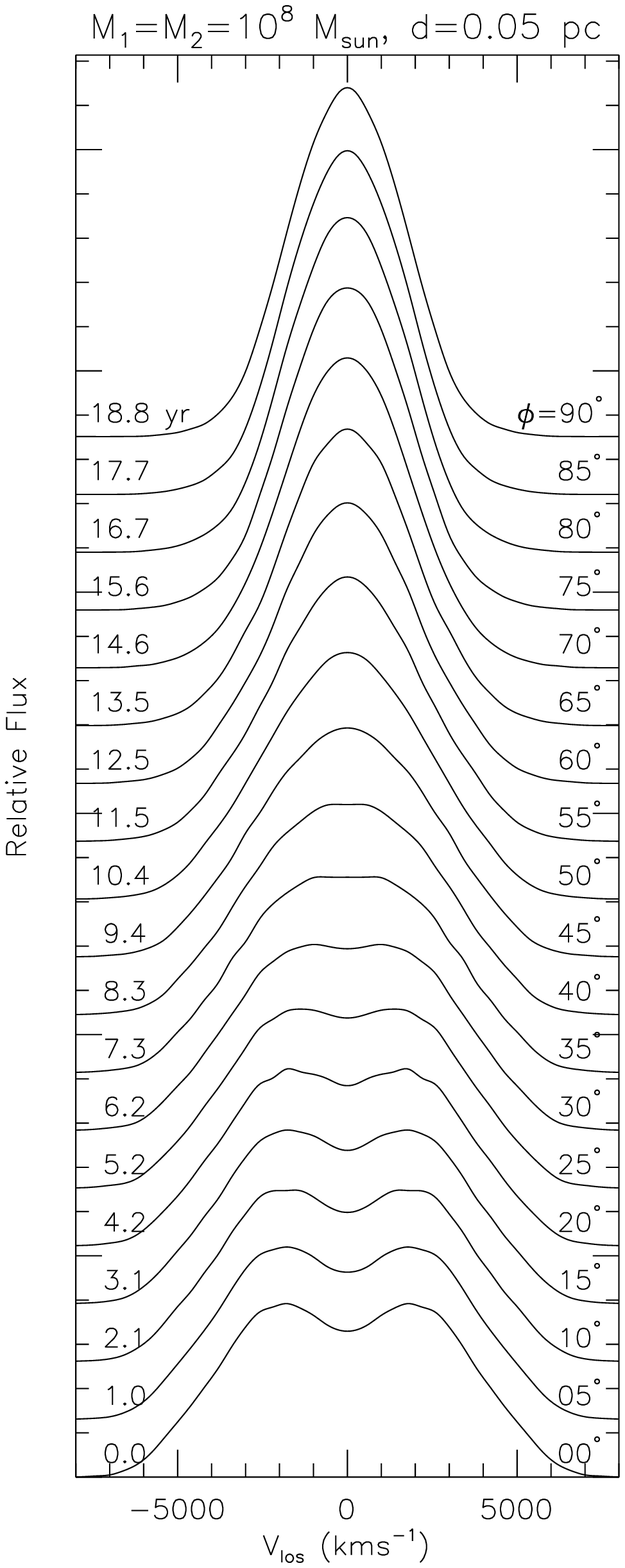}
    \includegraphics[width=0.45\textwidth]{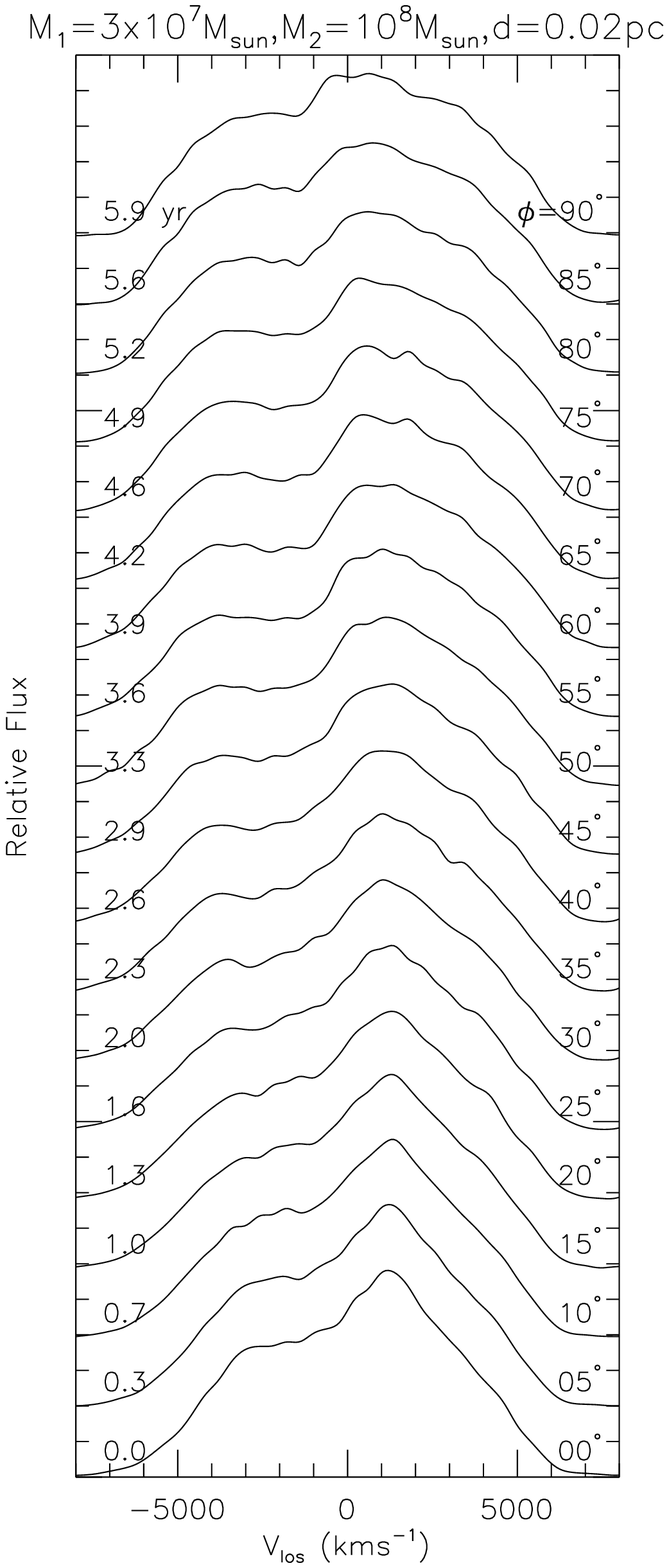}
    \caption{Time series of the radial velocity drifts in the line profile for an edge-on binary SMBH and for
     $1/4$ of the binary orbital period.
    {\em Left}: a binary with $M_1=M_2=10^8\ M_\odot$, $d=0.05$ pc, and $P\approx 75\ $yr (see Fig.\ \ref{fig:M1d8_d0d05}). The two BLRs are distinct in
    this case, and the drifts in
    radial velocities of the two peaks are apparent. {\em Right}: a binary
    with $M_1=3\times 10^7\ M_\odot$, $M_2=10^8\ M_\odot$, $d=0.02$ pc and $P\approx 23.5\ $yr (see Fig.\ \ref{fig:M3d7_M1d8_d0d02}). The two BLRs are no longer
    distinct in this case, and there are no coherent drifts in the radial velocities of the two peaks.}
    \label{fig:time_series}
\end{figure*}

\begin{figure*}
  \centering
    \includegraphics[width=0.48\textwidth]{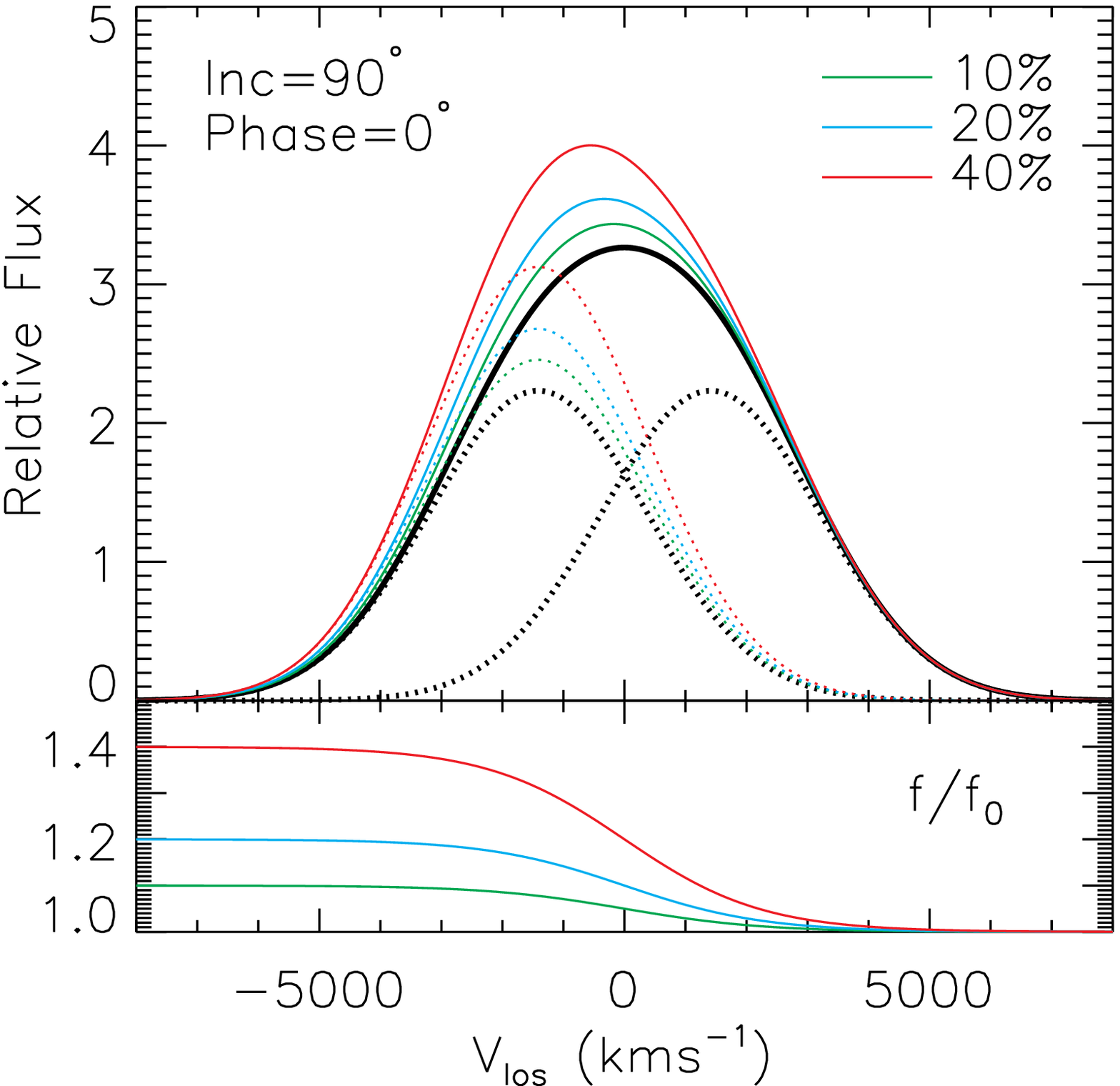}
    \includegraphics[width=0.48\textwidth]{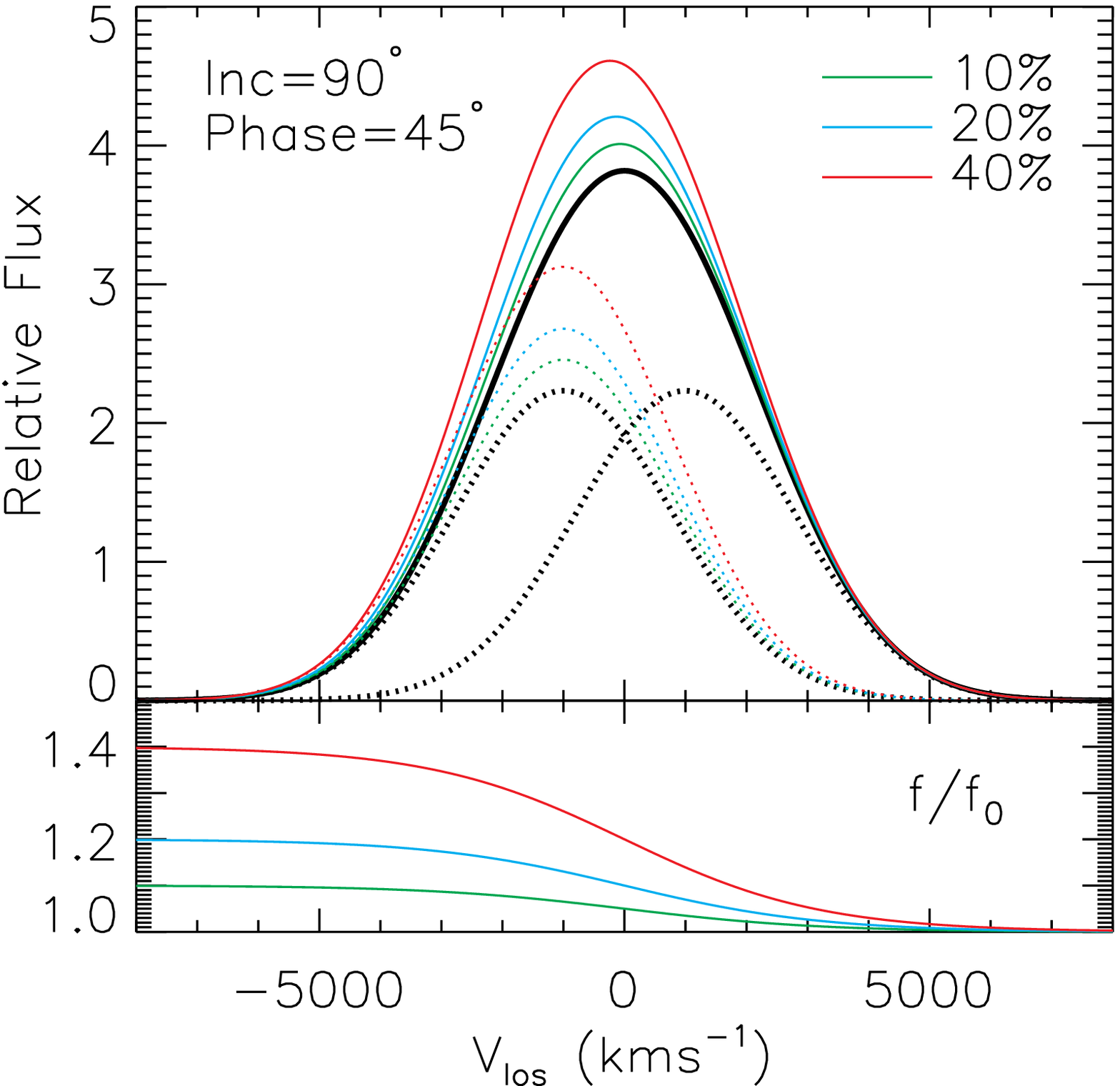}
    \caption{Responses of broad line profiles to continuum variations of one BH for
    the wide separation case. Shown here is an example of two $10^8\,M_\odot$ BHs with a separation
    $d=0.1$ pc, emitting at $\lambda_{\rm Edd}=0.1$ and in an edge-on view, with orbital
    phases $0$ and $\pi/4$.
    Solid lines show the overall line profile and dotted lines show individual
    components. The black lines show the original line profile, and the colored lines show the final profile after the reverberation completes. The bottom panels show the velocity-resolved flux
    changes.
    Even though the overall line profile is only mildly asymmetric, the fact that only
    one BLR is reverberating is seen once we decompose the broad line profile into two
    Gaussian components.}
    \label{fig:temp_vari_wide}
\end{figure*}

\begin{figure*}
  \centering
    \includegraphics[width=0.48\textwidth]{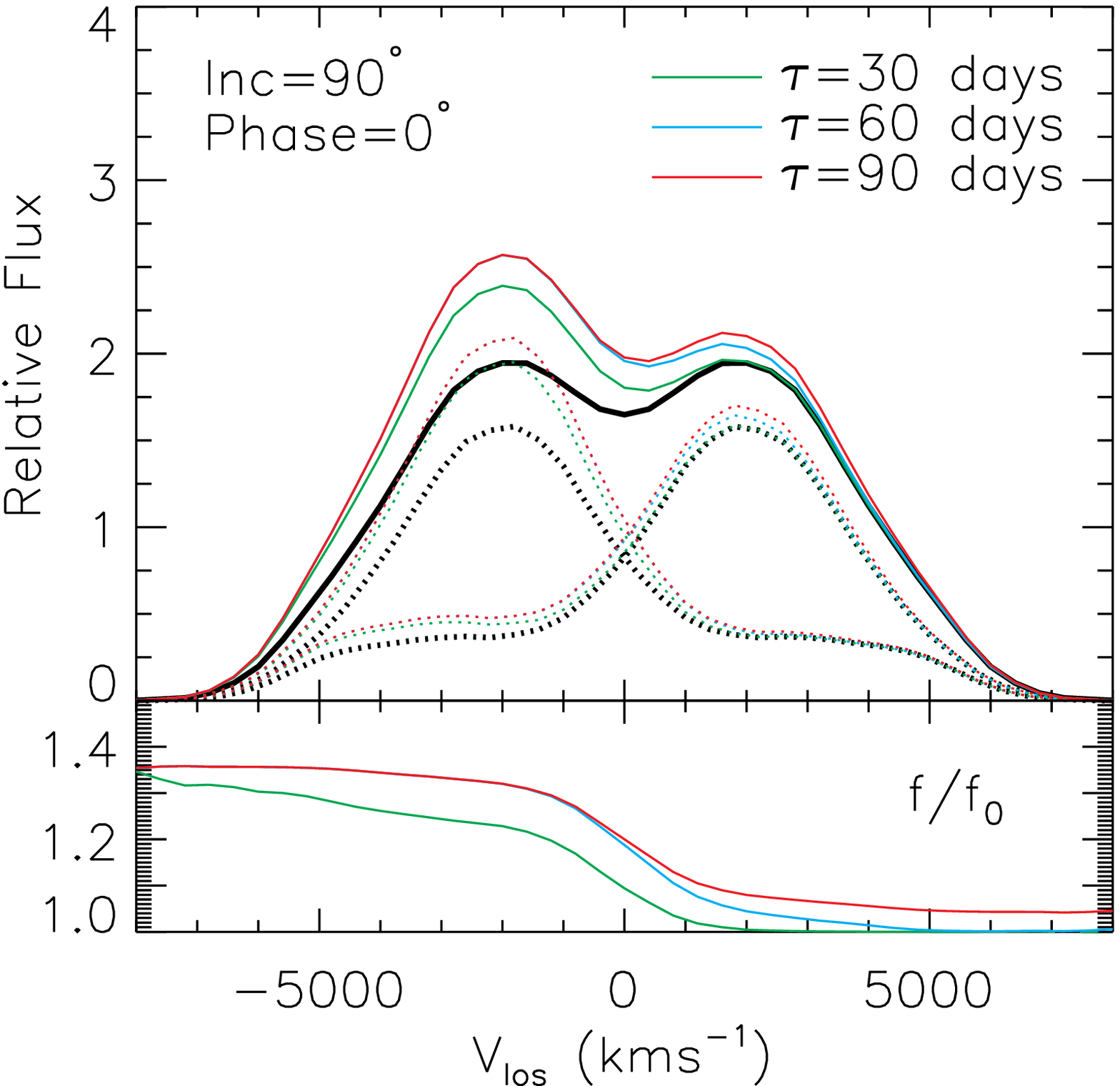}
    \includegraphics[width=0.48\textwidth]{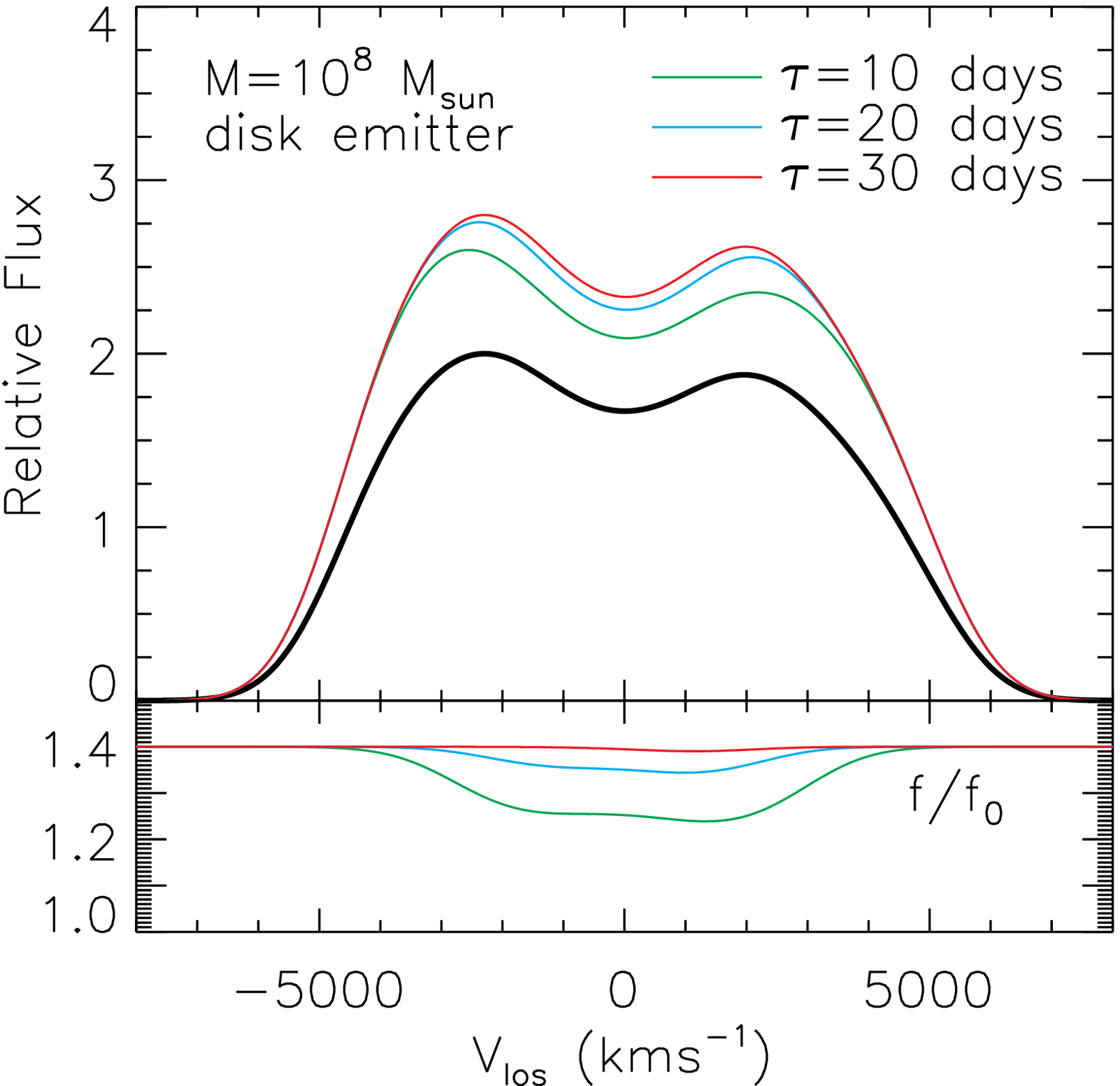}
    \caption{{\em Left:} Responses of broad line profiles to continuum variations of one BH for
    a binary with an intermediate separation. Shown here is an example of two $10^8\,M_\odot$ BHs with a separation
    $d=0.05$ pc, emitting
    at $\lambda_{\rm Edd}=0.1$ and in an edge-on view, with orbital phase $\phi=0$.
    We assume a $40\%$ increase in the continuum of one BH and derive the line response at different
    late times. Solid lines show the overall line profile and dotted lines show individual
    components. The black lines show the original line profile. The line response is completed at the last shown time epoch ($\tau=90$ days). The continuum variability of one BH also has effects on the BLR clouds of the other
    BH, which are delayed compared to the response of its own BLR clouds and are less prominent. {\em Right:} Reverberation mapping for a disk emitter around a single $10^8\ M_\odot$ BH. The amplitude of the continuum
    increase is $40\%$. The black line shows the original line profile. The disk model is the same as in the middle panel of Fig.\ \ref{fig:M1d8_d0d05}. The line
    response is completed at the last shown time epoch ($\tau=30$ days).}
    \label{fig:temp_vari_inter}
\end{figure*}

\subsection{Radial Velocity Drifts}\label{sec:dist1}
A definitive signature of a binary SMBH is the time drift in the  radial velocities of the decomposed two components, as resulting from the orbital motion of the
two BHs. Unfortunately, the typical orbital time is usually much
longer than a few years, and in order to detect radial velocity drifts of the double peaks the two BLRs must be distinct. From the previous sections we know that the optimal configuration to detect such a binary is when the two BLRs are just touching each other, such that the two BLRs are still mostly distinct while at the same time the velocity splitting of the two components is larger than the line width. In the case of two equal mass BHs, substituting $d=2R_{\rm BLR}$ in Eq.\
(\ref{eqn:bin_p}) where $R_{\rm BLR}$ is given by Eq.\ (\ref{eqn:BLR_size}), we have $P_{\rm opt}\sim 62(M_\bullet/10^8\
M_\odot)^{1/4}$ yr. This means it is less challenging to detect
radial velocity changes in low mass SMBH binaries by spectral
monitoring. In the case of two $M_\bullet=10^6\ M_\odot$, the time
span between double-peaked and single-peaked profile is only
$P_{\rm opt}/4\sim 5$ yr. The limitation of low-mass binaries is
that they cannot be easily observed out to high redshifts due to
their relatively low luminosities. Nevertheless, these low-mass
systems provide good test cases in the nearby Universe.

Figure \ref{fig:time_series} shows changes in the overall line
profile due to the orbital motion of the binary for an
intermediate separation case (the example in Fig.\
\ref{fig:M1d8_d0d05}) and for a close separation case (the example
in Fig.\ \ref{fig:M3d7_M1d8_d0d02}). While the radial velocity
drifts of the two peaks in the intermediate separation case are
apparent, there are no coherent drifts in the radial velocities of
the two peaks in the close separation case when the two BLRs are
no longer distinct (see the upper panel of Fig.\
\ref{fig:M3d7_M1d8_d0d02}). The non-detection of coherent radial velocity drifts in some of the double-peaked broad line AGNs may then rule out the existence of two distinct BLRs \citep[e.g.,][]{Eracleous_etal_1997,Gezari_etal_2007}, but cannot rule out the possibility of a close SMBH binary surrounded by a circumbinary BLR, as pointed out in \citet[][]{Eracleous_etal_1997}. Therefore additional tests are required to distinguish the binary and disk emitter scenarios in these cases\footnote{In addition to the reverberation mapping method discussed in \S \ref{sec:dist2}, there could be variations due to the orbital motion (Doppler effect) of the continuum emitting region of the accretion disk around each BH (Kocsis \& Loeb 2010, in preparation).} (see below).

\subsection{Reverberation Mapping}\label{sec:dist2}

A better way to distinguish the binary scenario and the disk
emitter scenario is reverberation mapping\footnote{In
principle, one could argue that independent variations in the two
velocity components of the broad line may be sufficient to
distinguish the binary interpretation from disk emitters
\citep[e.g.,][]{Gaskell_2009}. However, reverberation mapping
gives a cleaner signature since the variation in the broad line is
known to be caused by the variation in the continuum and not by changes
in the structure of the BLR.}
\citep[e.g.,][]{Peterson_etal_1987,Gaskell_2009}. Due to the
different geometries of the BLR in the two scenarios, the response
of the line to the variations in the continuum follows different
patterns.

In the case of wide separation binaries, each BLR is illuminated
by its own BH, and therefore only responds to the luminosity
variations of its BH. We demonstrate this case with the example of
two $10^8\ M_\odot$ BHs with a separation $0.1$ pc. The orbital
time of the binary is $\sim 200$ yr, while the typical variability
timescale for these BH masses is a few months
\citep[e.g.,][]{Peterson_1997}. Once the continuum of one of the
BHs varies, it takes $\sim 1$ month for the associated BLR
emission to vary, and the orbital motion of the binary during this
light travel time is negligible. The effects of continuum variations of one BH on the overall broad line profile are shown in
Fig.\ \ref{fig:temp_vari_wide} for several variability levels and
two orbital phases, after the broad line has completed its reverberation ($t\ga 35\,$days). Since the binary separation is large, the
broad line is not double-peaked. But the line response introduces
an asymmetry that correlates with the amplitude of the continuum
variation. This asymmetry is more prominent at smaller orbital
phase angles where the two velocity components overlap less in the spectrum.

In the case of intermediate separation binaries, the two BHs are
close enough such that part of the BLRs are illuminated by both
BHs. Consider the example of two $10^8\ M_\odot$ BHs with a
separation $0.05$ pc, i.e., about twice the size of a single BLR.
In this case, we use the configuration of BLR clouds derived in
our orbital integrations to compute the line profile changes due
to continuum changes of one BH. The orbital time of the binary is
$\sim 75$ yr, hence the orbital motion is negligible compared to
the light travel time to the BLR, which is a few months. However, due
to the proximity of the two BHs, the continuum variation in one of
the BHs also affects the other BLR, although the effect is delayed
and less prominent than those on its own BLR. The left panel in Figure
\ref{fig:temp_vari_inter} demonstrates these effects
for an edge-on view and orbital phase $\phi=0^\circ$. For comparison, the right panel shows the expected
line responses to continuum variations for the same disk emitter
model shown in Fig.\ \ref{fig:M1d8_d0d05} at various times, where the
line response is complete at the last epoch. Although the
high-velocity wing responses more quickly than the central part of
the line (because the former emission originates closer to the
BH), the two peaks change simultaneously. Even though initially
the broad line profiles are similar in the binary case and in
the disk emitter case, their subsequent line responses are
different: ii) the disk emitter completes the response of the line
much faster than the binary due to the proximity of the disk to
the central BH; ii) both peaks are varying in proportion to each
other in the disk emitter case, while in the binary case one peak
has larger amplitude changes than the other; and iii) the two peaks reverberate simultaneously in the disk emitter case, whereas in the binary case one peak reverberates faster than the other.

In the case of close separation binaries where the two BLRs are
already merged, the behavior of line response to continuum
variations of one of the BHs is similar to that of the
intermediate separation binary, but the relative increase in the
two peaks is less discrepant than in the intermediate separation
binary. Nevertheless, reverberation mapping will be a useful test
here because the tests on radial velocity drifts are inconclusive,
as discussed in \S \ref{sec:dist1}.

\section{Discussion}\label{sec:disc}

\begin{figure}
  \centering
    \includegraphics[width=0.48\textwidth]{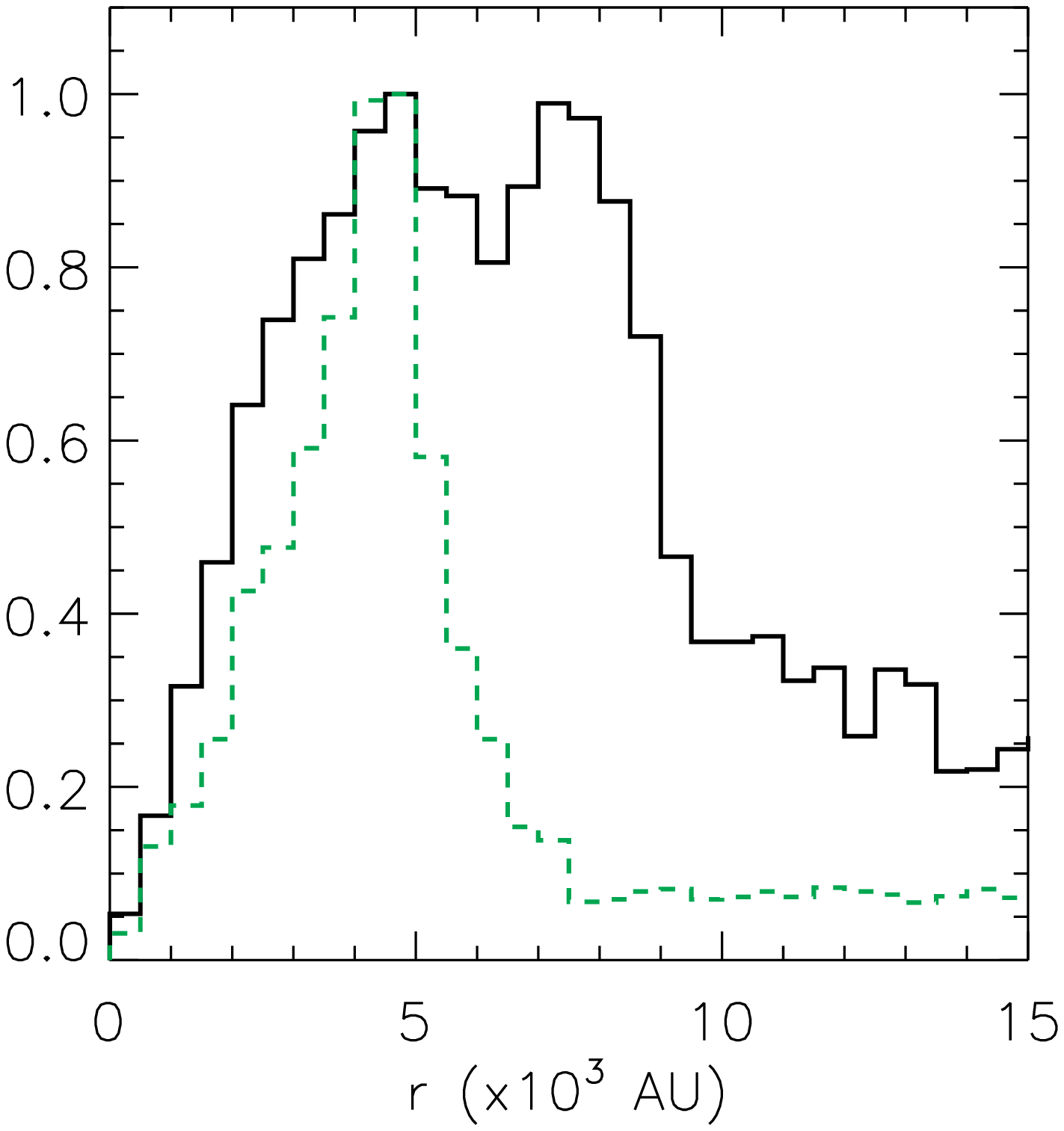}
    \includegraphics[width=0.48\textwidth]{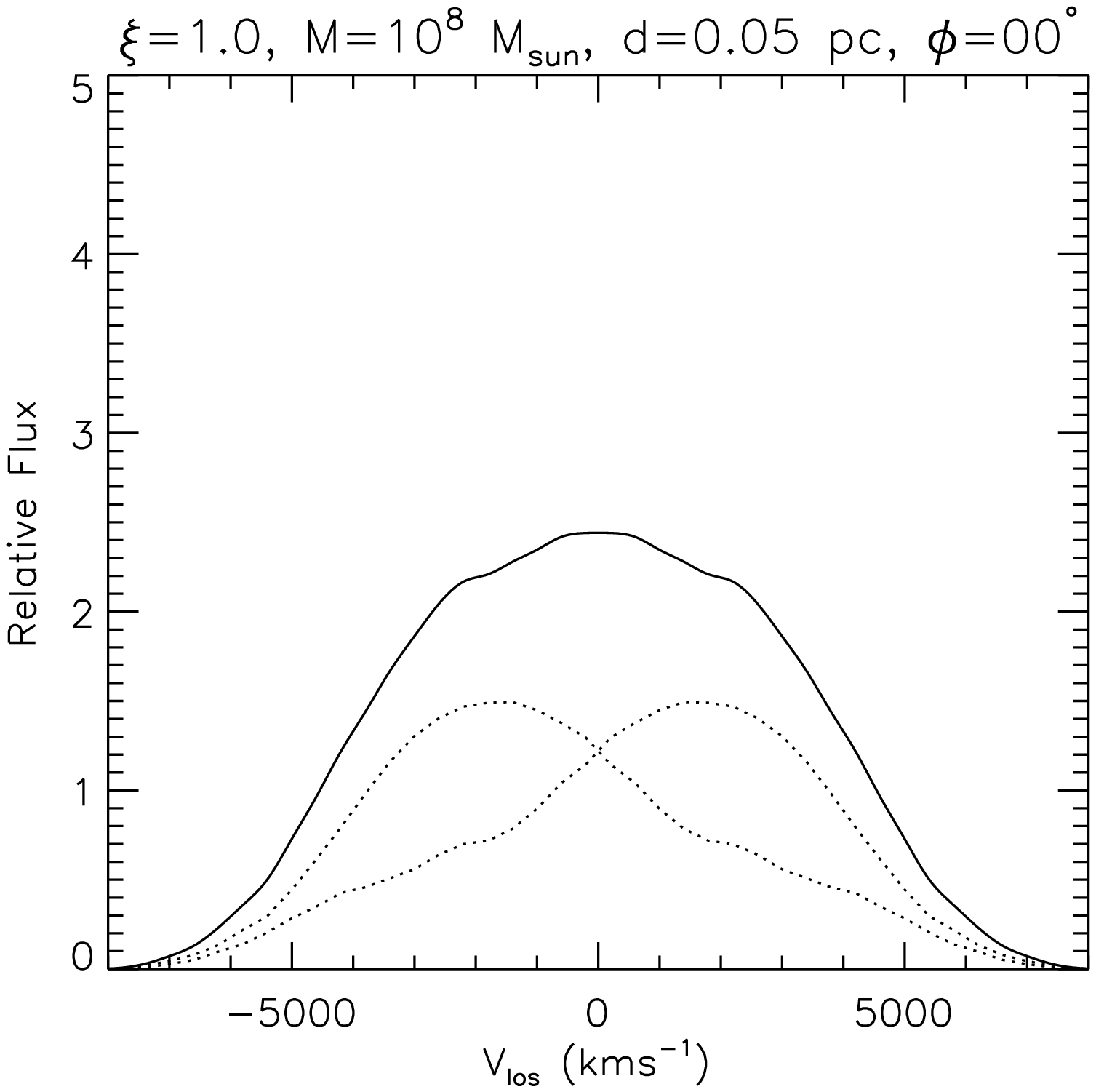}
    \caption{Results for an extended cloud distribution around a single $M=10^8\,M_\odot$ BH. {\em Upper:} Radial distributions of clouds around the single BH as initial conditions for the numerical orbit integrations. The green dashed line is the fiducial cloud distribution described in \S \ref{sec:model}, which peaks around the characteristic radius given by Eqn.\ (\ref{eqn:BLR_size}). The black line is a more extended cloud distribution (see \S \ref{subsec:disc1} for details). {\em Bottom:} The resulting line profile for the extended cloud distribution and for an equal-mass binary ($M_{\rm tot}=2\times 10^8\,M_\odot$) with a binary separation $d=0.05\,$pc at phase angle $\phi=0^\circ$. The two line components are much less distinct than in the previous case (e.g., Fig.\ \ref{fig:M1d8_d0d05}) because a larger fraction of the outmost BLR clouds are on circumbinary orbits. This tends to dilute the distinction between the two components from BLR clouds orbiting around individual BHs.}
    \label{fig:alter_BLR}
\end{figure}

\begin{figure}
  \centering
    \includegraphics[width=0.48\textwidth]{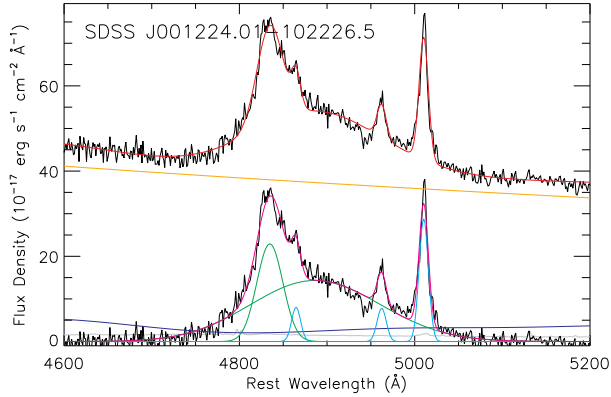}
    \caption{An example of double-peaked broad line AGNs from SDSS. The black lines show
    the original SDSS spectrum (upper) and the continuum/iron flux subtracted spectrum
    (bottom). The red and magenta lines show the overall model fits. The orange and blue lines
    are the power-law continuum and iron template fits.
    The three Gaussian components in cyan are the narrow lines \hbeta, \OIIIa\ and \OIIIb,
    fixed to have the same redshifts and line widths. The two green Gaussian components are for
    the double-peaked broad line profile, which are separated by $\sim 3000\ {\rm kms^{-1}}$, and
    have FWHMs $\sim 2200\ {\rm kms^{-1}}$ and $\sim 10000\ {\rm kms^{-1}}$ for the blueshifted
    and redshifted components respectively. This object and many others will be good candidates for
    spectral monitoring and reverberation mapping programs.}
    \label{fig:dp_examp}
\end{figure}

\subsection{Complications in the Realistic Situation}\label{subsec:disc1}
Our simple prescription for the BLR of a SMBH binary is by no
means realistic, especially for the closest separation cases
studied here, where there are no longer two distinct BLRs. The BLR
models we adopted are close to models in which the orbits are
random, and the dynamics is dominated by the gravitational
potential from both BHs. The novelty of our approach is to include
the effects of the two BHs in terms of both the clouds dynamics
and illumination. Our treatment is more quantitative than earlier
qualitative arguments that the double-peaked components are from
individual BLRs \citep[e.g.,][and references
therein]{Gaskell_2009}.

In our BLR model for single BHs we assumed a rather simplistic thin shell distribution of clouds. To check the sensitivity of our results to this assumption, we examine the effects of a more extended cloud (not all line-emitting clouds) distribution for single BHs. The upper panel of Fig.\ \ref{fig:alter_BLR} shows an example for the distribution of clouds around a $10^8\,M_\odot$ BH, where the clouds were initially populated between 0.5 and 2 times the characteristic BLR size with a power-law number density $n(r)\propto r^{-1}$ and then relaxed for 30 years using their Keplerian orbits. The initial random velocities of each cloud are assigned using the same scheme described in \S\ref{sec:model}. The starting configuration for the numerical orbit integration is a more extended distribution of clouds compared to our fiducial model. We integrate an equal-mass binary system ($M_{\rm tot}=2\times 10^8\,M_\odot$) using the new single BLR model and compute emission line profiles in the same way as in \S \ref{sec:model}. The bottom panel of Figure \ref{fig:alter_BLR} shows the resulting line profile at phase angle $\phi=0^\circ$ and with a binary separation $d=0.05\,$pc. Compared with our previous result, i.e., the middle panel of Fig.\ \ref{fig:M1d8_d0d05}, the double-peaked feature is much less prominent. This is expected because now the two cloud regions overlap more than in the previous case and more clouds become circumbinary, diluting the distinction between the two BLRs. In practice, the two emission line components will not have equal strength, so some asymmetry is expected in the overall line profile, similar to those shown in Fig.\ \ref{fig:M3d7_M1d8_d0d05}. Secular changes in the line profile due to the orbital motion of the binary, or velocity resolved reverberations of the blue and red wings of the line, can still be monitored even though the two peaks are blended with each other.

Our toy models confirm the feasibility of using spectral
monitoring (for radial velocity drifts) and reverberation mapping
(for line responses) to disentangle SMBH binaries from disk
emitters. These observations can also be achieved with less
expensive narrow-band filters. However, there are some practical
difficulties with these techniques. The spectral monitoring for
radial velocity changes works best for binaries with $d\approx R_{\rm
BLR,1}+R_{\rm BLR,2}$. At larger binary separations, the two broad
components will blend with each other in the spectrum, making it less likely to be
flagged as a binary candidate, and the orbital period is too long
to be detectable. At smaller separations, the BLRs are no longer
distinct and the velocity structure becomes more complex with no
coherent radial velocity drifts in the peak locations. Thus, the
spectral monitoring is suitable for identifying low mass SMBH binaries ($M\sim 10^6\ M_\odot$). On the other hand, reverberation mapping is a powerful tool for distinguishing a SMBH binary from a disk emitter, but more complex BLR geometries and kinematics (such as those involving inflows or outflows) will certainly complicate the
situation \citep[e.g.,][]{Sergeev_etal_1999,Denney_etal_2009}.

\subsection{Case Studies: 3C 390.3, SDSS J1536+0441 and others}

Although there are a few dedicated spectral monitoring programs
for double-peaked broad line objects \citep[e.g.,][and references
therein]{Gezari_etal_2007}, there is currently no reverberation
mapping program for a large sample of such objects. Among the
$\sim 40$ AGNs that are included in the reverberation mapping
sample \citep[][]{Peterson_etal_2004}, there are several objects
that clearly show double-peaked broad line features. In particular
3C 390.3 is a strong double-peak object with good reverberation
mapping data \citep[][]{Dietrich_etal_1998}. The time-ordered data
of this source show that the blueshifted and redshifted components
respond to the continuum variations almost simultaneously. Thus,
it is more likely that the double-peaked emission originates from
a disk rather than from two corotating BLRs in a binary system.

The quasar SDSS J1536+0441 was recently suspected to be a sub-pc binary
SMBH \citep[][]{Boroson_Lauer_2009} because of its double-peaked
broad \hbeta\ line in the SDSS spectrum. It has therefore received
much attention
\citep[e.g.,][]{Chornock_etal_2009,Gaskell_2009,Wrobel_Laor_2009,Decarli_etal_2009,Tang_etal_2009}.
The discovery of an additional redshifted component, most notably
in \halpha\
\citep[e.g.,][]{Chornock_etal_2009,Lauer_Boroson_2009}, favored a
disk emitter origin for the double peaks rather than a binary
SMBH. If we nevertheless assume this is a binary SMBH and use
constant Eddington ratios $\lambda_{\rm Edd}=0.1$ for both BHs,
and the FWHM values measured in \citet[][]{Boroson_Lauer_2009}, we
get from equation (\ref{eqn:FWHM}) $M_1=8\times 10^8\ M_\odot$ and
$M_2=2\times 10^7\ M_\odot$ for the red and blue systems,
respectively, similar to those reported in
\citet{Boroson_Lauer_2009}. We also derive BLR sizes
$R_1\approx0.063\ {\rm pc}$ and $R_2\approx0.01\ {\rm pc}$.
Substituting the BH masses and $v_{\rm los}=3500\ {\rm kms^{-1}}$
in equation (\ref{eqn:v_split}) we get $d=0.27[\sin I\sin(2\pi
t/P)]^2\ {\rm pc}$. However, in order to produce the comparable
strength of both components and hence two distinct peaks, it
requires that the smaller BH (blue component) is $\sim 40$ times
more efficient at producing the broad-line emission than the
larger BH. Furthermore, by comparing the spectrum taken $\sim 1$
yr after \citep[][]{Chornock_etal_2009} with the original SDSS
spectrum, it appears that both the blueshifted and the redshifted
components become slightly weaker whereas the central component
(which would be the classic BLR in the disk emitter scenario)
remains almost unchanged. This strengthens the association of a
disk emitter origin with the double-peaked feature.

There have been a significant number of double-peaked or highly asymmetric
broad line AGNs known in the literature, and large spectroscopic surveys such as
SDSS are providing many more
\citep[e.g.,][]{Strateva_etal_2003,Shen_etal_2010}. 
An example, SDSS J0012-1022\footnote{Strictly speaking, this is not a double-peaked profile, but an asymmetric profile with an extended red wing. Nevertheless it was included in the double-peaked Balmer line AGN sample in \citet{Strateva_etal_2003}.}, is shown in Fig.\
\ref{fig:dp_examp}, whose broad line profile can be well fitted by two components. Note that the velocity splitting of this object is smaller than the FWHM of the
redshifted component, and both broad components are reasonably well-fitted by a single Gaussian. If it is of a binary origin, its separation must be large enough such that the
two BLRs are mostly distinct. Even though
its profile is not obviously double-peaked, it still stands out as a promising binary candidate. It would be interesting to perform spectral monitoring and
reverberation mapping for a large statistical sample of such objects in order to determine their nature.

\subsection{The Frequency of Spectroscopic
Binaries}\label{sec:dist3}

There is currently some tension between theoretical expectations
and observations on the frequency of SMBH binaries: if major
mergers of gas-rich galaxies is the triggering mechanism of quasar
activity, then the expected binary fraction is very high; on the
other hand, the observed frequency of binary SMBHs is less than a
few percent beyond pc scales, and much lower on pc to sub-pc
scales. We can parameterize the ``observable'' binary fraction at
different stages as the product of several factors:
\begin{equation}\label{eqn:f_binary}
f_{\rm obs}=f_{\rm bin}\times f_{\rm active}\times f_{\rm
geo}\times f_{\rm tech}\times (\tau_{\rm phase}/\tau_{\rm QSO})\ ,
\end{equation}
where $f_{\rm bin}$ is the intrinsic binary fraction ($f_{\rm
bin}\approx 1$ in the merger hypothesis for quasar activity),
$f_{\rm active}$ is the probability of both BHs being active,
$f_{\rm geo}$ is the observable fraction due to orientational
effects (inclination and orbital phase), $f_{\rm tech}$ is the
observable fraction due to the specific technique used (i.e.,
spectroscopic methods or spatially resolved imaging) which depends
on the data quality of the observing program and folds in the
complications associated with the emission line region geometry,
$\tau_{\rm phase}$ is the time span during the specific stage of
binary evolution, and $\tau_{\rm QSO}$ is the lifetime of quasars.
All these quantities except for $\tau_{\rm QSO}$ are functions of
the evolutionary stage of the binary.

In this paper, we have focused on the broad line diagnosis
technique, which are for parsec to sub-parsec binaries before the
binary enters the gravitational wave dominated regime. Therefore,
we expect $f_{\rm active}\approx 1$ since in gas-rich mergers a
nuclear gas disk on pc to 100 pc scales may form and feed both BHs
\citep[e.g.,][]{Mayer_etal_2007}. For the effects of random
orientations and orbital phases it is reasonable to adopt $f_{\rm
geo}\la 0.25$, since nearly edge-on systems with radial motions
are most likely to be detected. $f_{\rm tech}$ is difficult to
quantify without a dedicated program and related Monte-Carlo
simulations, but it is unlikely that $f_{\rm tech}$ is close to
unity. Bearing in mind the large uncertainties, the time span
during this stage can be estimated as $\sim 10^5-10^7$ yr if gas
drag is the dominant mechanism that shrinks the binary orbit
\citep[e.g.,][]{Escala_etal_2005,Dotti_etal_2006,Mayer_etal_2007}.
The quasar lifetime is not very well constrained and typical
values are $\tau_{\rm QSO}\sim 10^7-10^8$ yr. Taken together, the
fraction of observable parsec to sub-parsec binaries based on
broad line diagnosis is less than a few percent. This is still
higher than the frequency of the known parsec to sub-parsec binary
candidates, but lower than the frequency of known
double-peaked broad line AGNs \citep[e.g.,][and references
therein]{Strateva_etal_2003,Gezari_etal_2007}. Of course, these
are very crude estimates, and measuring the actual observed
frequency will serve an important role of testing theoretical models of binary formation and evolution \citep[e.g.,][]{Volonteri_etal_2009}.

\acknowledgments

We thank the anonymous referee for constructive comments that improved the manuscript. YS acknowledges support from a Clay Postdoctoral Fellowship
through the Smithsonian Astrophysical Observatory (SAO). This work
was supported in part by NSF grant AST-0907890 and NASA grants NNX08AL43G and NNA09DB30A (AL).


\label{lastpage}
\end{document}